\documentclass[12pt]{article}
\usepackage{a4p}
\usepackage{epsfig}
\usepackage{amssymb}
\usepackage{amsmath}
\usepackage{cite}

\def\opalackerstaff{OPAL Collaboration, K.\ Ackerstaff \etal}
\def\opalabbiendi{OPAL Collaboration, G.\ Abbiendi \etal}
\def\opalalexander{OPAL Collaboration, G.\ Alexander \etal}
\def\opalakers{OPAL Collaboration, R.\ Akers \etal}

\def\opalakrawy{OPAL Collaboration, M.Z.\ Akrawy \etal}
\def\opalahmet{OPAL Collaboration, K.\ Ahmet \etal}
\newcommand{\LEP}{\mbox{LEP}}
\newcommand{\LEPI}{\mbox{LEP1}}
\newcommand{\OPAL}{\mbox{OPAL}}
\newcommand{\Opal}{\mbox{OPAL}}

\newcommand{\WW}{\ensuremath{\mathrm{W}^+\mathrm{W}^-}}
\newcommand{\epem}{\ensuremath{\mathrm{e}^+\mathrm{e}^-}}
\newcommand{\roots}{\ensuremath{\sqrt{s}}}

\newcommand{\Mw}{\ensuremath{M_{\mathrm{W}}}}

\newcommand{\stat}{\mathrm{(stat.)}}
\newcommand{\syst}{\mathrm{(syst.)}}
\newcommand{\theory}{\mathrm{(theory)}}
\newcommand{\Br}{\ensuremath{\mathrm{Br}}}
\newcommand{\Wtomn}{\mbox{$\mathrm{W}\rightarrow\mnu$}}
\newcommand{\Wtoen}{\mbox{$\mathrm{W}\rightarrow\enu$}}
\newcommand{\Wtotn}{\mbox{$\mathrm{W}\rightarrow\tnu$}}

\newcommand{\Wtoqq}{\mbox{$\mathrm{W}\rightarrow\qq$}}
\newcommand{\sigWW} {\ensuremath{\sigma_{_{\mathrm{WW}}}}}
\newcommand{\sigWWSM} {\ensuremath{\sigma_{_{\mathrm{WW}}}^{_{\mathrm{SM}}}}}
\newcommand{\costw} {\ensuremath{\cos\theta_{\mathrm{W}^-}}}
\newcommand{\nue}   {\ensuremath{\nu_{e}}}
\newcommand{\nuebar}{\ensuremath{\overline{\nu}_{e}}}
\newcommand{\numu}  {\ensuremath{\nu_{\mu}}}
\newcommand{\nutau} {\ensuremath{\nu_{\tau}}}

\newcommand{\enu}   {\ensuremath{\mathrm{e}\nue}}
\newcommand{\mnu}   {\ensuremath{\mu\numu}}
\newcommand{\tnu}   {\ensuremath{\tau\nutau}}

\newcommand{\qq}    {\ensuremath{\mathrm{qq}}}
\newcommand{\lplm}  {\ensuremath{\ell^+\ell^-}}

\newcommand{\nunu}  {\ensuremath{\nu\overline{\nu}}}

\newcommand{\ffbar} {\ensuremath{f\overline{f}}}
\newcommand{\mpmm} {\ensuremath{{\mu}^+{\mu}^-}}
\newcommand{\tptm} {\ensuremath{{\tau}^+{\tau}^-}}

\newcommand{\enens}{\ensuremath{\enu\enu}}
\newcommand{\mnmns}{\ensuremath{\mnu\mnu}}
\newcommand{\tntns}{\ensuremath{\tnu\tnu}}
\newcommand{\enmns}{\ensuremath{\enu\mnu}}
\newcommand{\entns}{\ensuremath{\enu\tnu}}
\newcommand{\mntns}{\ensuremath{\mnu\tnu}}
\newcommand{\lnln}{\ensuremath{\ell\nu\ell\nu}}
\newcommand{\lv}{\ensuremath{\ell\nu}}
\newcommand{\lvlv}{\ensuremath{\ell\nu\ell\nu}}
\newcommand{\qqen}{\ensuremath{\qq\enu}}
\newcommand{\qqmn}{\ensuremath{\qq\mnu}}
\newcommand{\qqtn}{\ensuremath{\qq\tnu}}
\newcommand{\qqln}{\ensuremath{\qq\ell\nu}}
\newcommand{\qqlv}{\ensuremath{\qq\ell\nu}}
\newcommand{\qqqq}{\ensuremath{\qq\qq}}
\newcommand{\lnlns}{\ensuremath{\ell\nu\ell\nu}}
\newcommand{\qqlns}{\ensuremath{\mathrm{qq}\ell\nu}}

\newcommand{\gW}{\ensuremath{g_\mathrm{W}}}

\newcommand{\WWqqen}{\mbox{\WW$\rightarrow$ \qqen}}
\newcommand{\WWqqmn}{\mbox{\WW$\rightarrow$ \qqmn}}
\newcommand{\WWqqtn}{\mbox{\WW$\rightarrow$ \qqtn}}
\newcommand{\WWlnln}{\mbox{\WW$\rightarrow$ \lnln}}
\newcommand{\WWqqln}{\mbox{\WW$\rightarrow$ \qqln}}
\newcommand{\WWqqqq}{\mbox{\WW$\rightarrow$ \qqqq}}

\newcommand{\qqnn}{\ensuremath{\qq\nunu}}
\newcommand{\qqll}{\ensuremath{\qq\lplm}}
\newcommand{\qqee}{\ensuremath{\qq\epem}}
\newcommand{\qqmm}{\ensuremath{\qq\mpmm}}
\newcommand{\ZZqqqq}{\ensuremath{\mathrm{ZZ}\rightarrow\qq\qq}}

\newcommand{\eeff}{\ensuremath{\epem\ffbar}}

\newcommand{\eetoqq}{\ensuremath{\epem\rightarrow\qq}}
\newcommand{\eetoZZ}{\ensuremath{\epem\rightarrow\Zz\Zz}}

\newcommand{\Zz}       {\ensuremath{{\mathrm{Z}}}}
\newcommand{\ZzZz}{\ensuremath{\Zz\Zz}}

\newcommand{\Wenu}     {\ensuremath{\mathrm{W}\enu}}
\newcommand{\Zee}      {\ensuremath{\mathrm{Z}\epem}}

\newcommand{\qqgg}     {\ensuremath{\qq \mathrm{gg}}}
\newcommand{\qqev}{\ensuremath{\qq\mathrm{e}\nu}}
\newcommand{\qqmv}{\ensuremath{\qq\mu\nu}}
\newcommand{\qqtv}{\ensuremath{\qq\tau\nu}}
\newcommand{\llnunu}{\ell\nu\ell\nu}

\newcommand{\CC}{\mbox{{\sc CC03}}}

\newcommand{\JETSET}{\mbox{J{\sc etset}}}
\newcommand{\KORALW}{\mbox{K{\sc oral}W}}

\newcommand{\EXCALIBUR}{\mbox{E{\sc xcalibur}}}
\newcommand{\GRC}{\mbox{grc4f}}
\newcommand{\PYTHIA}{\mbox{P{\sc ythia}}}
\newcommand{\ARIADNE}{\mbox{A{\sc riadne}}}
\newcommand{\BHWIDE}{\mbox{B{\sc hwide}}}
\newcommand{\KANDY}{\mbox{K{\sc and}}Y}
\newcommand{\PHOJET}{\mbox{P{\sc hojet}}}
\newcommand{\PHOTOS}{\mbox{P{\sc hotos}}}
\newcommand{\BDK}{\mbox{B{\sc dk}}}
\newcommand{\HERWIG}{\mbox{H{\sc erwig}}}
\newcommand{\KK}{\mbox{KK2f}}
\newcommand{\YFSWW}{\mbox{Y{\sc fs}WW}}
\newcommand{\RACWW}{\mbox{R{\sc acoon}WW}}

\newcommand{\WQCD}{\ensuremath{W_{420}}}
\newcommand{\WCC}{\ensuremath{W_{\mathrm{CC03}}}}
\newcommand{\Vij} {\mbox{$|{V}_{ij}|$}}
\newcommand{\Vud} {\mbox{$|{V}_{\mathrm{ud}}|$}}
\newcommand{\Vus} {\mbox{$|{V}_{\mathrm{us}}|$}}
\newcommand{\Vcd} {\mbox{$|{V}_{\mathrm{cd}}|$}}
\newcommand{\Vcb} {\mbox{$|{V}_{\mathrm{cb}}|$}}
\newcommand{\Vub} {\mbox{$|{V}_{\mathrm{ub}}|$}}
\newcommand{\Vcs} {\mbox{$|{V}_{\mathrm{cs}}|$}}
\newcommand{\pz}{\mbox{$p_{\mathrm{z}}^{\mathrm{miss}}$}}
\newcommand{\pT}{\mbox{$p_{\mathrm{T}}$}}
\newcommand{\xT}{\mbox{$x_{\mathrm{T}}$}}
\newcommand{\acop}{\mbox{$\phi_{\mathrm{acop}}$}}
\newcommand{\ptaxic}{\mbox{$a_{\mathrm{T}}^{\mathrm{miss}}$}}
\newcommand{\aT}{\mbox{$a_{\mathrm{T}}^{\mathrm{miss}}$}}
\newcommand{\athet}{\mbox{$\theta_{\mathrm{a}}^{\mathrm{miss}}$}}

\newcommand{\Sevev}{\mbox{e$\nu$e$\nu$}}
\newcommand{\Sevmv}{\mbox{e$\nu\mu\nu$}}
\newcommand{\Sevtv}{\mbox{e$\nu\tau\nu$}}
\newcommand{\Smvmv}{\mbox{$\mu\nu\mu\nu$}}
\newcommand{\Smvtv}{\mbox{$\mu\nu\tau\nu$}}
\newcommand{\Stvtv}{\mbox{$\tau\nu\tau\nu$}}
\newcommand{\Sqqev}{\mbox{\qq e$\nu$}}
\newcommand{\Sqqmv}{\mbox{\qq$\mu\nu$}}
\newcommand{\Sqqtv}{\mbox{\qq$\tau\nu$}}
\newcommand{\Sqqqq}{\mbox{\qq\qq}}
\newcommand{\Slvlv}{\mbox{$\ell\nu\ell\nu$}}
\newcommand{\Sqqlv}{\mbox{{\qq}$\ell\nu$}}
\newcommand{\Blvlv}{\mbox{$\ell\nu\ell\nu$}}

\newcommand{\Bqqlv}{\mbox{\qq$\ell\nu$}}

\newcommand{\Bqqqq}{\mbox{\qq\qq}}
\newcommand{\Bqqvv}{\mbox{\qq$\nu\nu$}}
\newcommand{\Bqqll}{\mbox{\qq$\ell\ell$}}
\newcommand{\Bllll}{\mbox{$\ell\ell\ell\ell$}}

\newcommand{\Bqq}{\mbox{\qq}}
\newcommand{\Bll}{\mbox{$\ell\ell$}}
\newcommand{\BeeX}{\mbox{\epem X}}
\newcommand{\eeX}{\mbox{$\epem\rightarrow\epem X$}}
\newcommand{\Gevev}{\mbox{e$\nu$e$\nu$}}
\newcommand{\Gevmv}{\mbox{e$\nu\mu\nu$}}
\newcommand{\Gevtv}{\mbox{e$\nu\tau\nu$}}
\newcommand{\Gmvmv}{\mbox{$\mu\nu\mu\nu$}}
\newcommand{\Gmvtv}{\mbox{$\mu\nu\tau\nu$}}
\newcommand{\Gtvtv}{\mbox{$\tau\nu\tau\nu$}}
\newcommand{\Gqqev}{\mbox{{\qq}e$\nu$}}

\newcommand{\Gqqmv}{\mbox{\qq$\mu\nu$}}
\newcommand{\Gqqtv}{\mbox{\qq$\tau\nu$}}
\newcommand{\Gqqqq}{\mbox{\qq\qq}}

\newcommand{\GeV}{\mbox{$\mathrm{GeV}$}}

\newcommand{\PLB}[3]  {Phys.\ Lett.\ \textbf{B#1} (#2) #3}
\newcommand{\ZPC}[3]  {Z.\ Phys.\ \textbf{C#1} (#2) #3}
\newcommand{\EPC}[3]  {Eur.\ Phys.\ J.\ \textbf{C#1} (#2) #3}
\newcommand{\NIMA}[3] {Nucl.\ Instr.\ and Meth.\ \textbf{A#1} (#2) #3}

\newcommand{\PRL}[3]  {Phys.\ Rev.\ Lett.\ \textbf{#1} (#2) #3}
\newcommand{\PRD}[3]  {Phys.\ Rev.\ \textbf{D#1} (#2) #3}
\newcommand{\NPB}[3]  {Nucl.\ Phys.\ \textbf{B#1} (#2) #3}

\newcommand{\JHEP}[3] {JHEP \textbf{#1} (#2) #3}
\newcommand{\CPC}[3]  {Comput.\ Phys.\ Commun.\ \textbf{#1} (#2) #3}
\newcommand{\CIC}[3]  {Comp.\ in Phys.\  \textbf{#1} (#2) #3}
\newcommand{\lumi}{701.1}
\newcommand{\nsel}{11693}
\newcommand{\totlumi}{\mbox{$\lumi\,\mathrm{pb}^{-1}$}}
\newcommand{\totlumiwitherror}{\mbox{$\lumi\pm 2.1\,\mathrm{pb}^{-1}$}}
\newcommand{\SMbrlv}{\ensuremath{10.83}\,\%}
\newcommand{\SMbrqq}{\ensuremath{67.51}\,\%}

\newcommand{\brevresult}{\ensuremath{10.71\pm0.25 \stat \pm0.11 \syst}\,\%}
\newcommand{\brmvresult}{\ensuremath{10.78\pm0.24 \stat \pm0.10 \syst}\,\%}
\newcommand{\brtvresult}{\ensuremath{11.14\pm0.31 \stat \pm0.17 \syst}\,\%}
\newcommand{\brqqresult}{\ensuremath{67.41\pm0.37 \stat \pm0.23 \syst}\,\%}
\newcommand{\sumvresult}{\ensuremath{1.993\pm0.033\stat\pm0.023\syst}}
\newcommand{\vcsresult}{\ensuremath{0.969\pm0.017\stat\pm0.012\syst}}
\newcommand{\gwratio}{\ensuremath{0.996\pm0.017\stat\pm0.011\syst}}
\newcommand{\rsmresult}{\ensuremath{1.002\pm0.011 \stat \pm0.007 \syst \pm 0.005 \theory}}

\def\etal{\mbox{{\it et al.}}}

\begin{document}
\begin{titlepage}
\begin{center}{\large   EUROPEAN ORGANIZATION FOR NUCLEAR RESEARCH
}\end{center}\bigskip
\begin{flushright}
       CERN-PH-EP/2007-027 \\  
       OPAL PR\,424 \\ 22 July 2007
\end{flushright}
\bigskip\bigskip\bigskip\bigskip\bigskip
\begin{center}\huge{\boldmath\bf Measurement of the  $\epem\rightarrow\WW$ cross section and W decay branching fractions  at LEP}
\end{center}\bigskip\bigskip
\begin{center}{\LARGE The OPAL Collaboration
}\end{center}\bigskip\bigskip
\bigskip\begin{center}{\large  Abstract}\end{center}

From a total data sample of \lumi~pb$^{-1}$ recorded with 
$\epem$ centre-of-mass energies of $\roots = 161-209$~GeV 
with the OPAL detector at LEP, \nsel\ W-pair 
candidate events are selected. These data are used to obtain
measurements of the W-pair production cross sections at
10 different centre-of-mass energies. The ratio of the
measured cross sections to the Standard Model expectation
is found to be:
$$ \mathrm{data}/\mathrm{SM} = \rsmresult,$$
where the uncertainties are statistical, experimental systematics and
theory systematics respectively. 
The data are used to determine the W boson branching fractions, which
are found to be consistent with lepton universality of the charged 
current interaction. Assuming lepton universality,  
the branching ratio to hadrons is determined to be \brqqresult,
from which the CKM matrix element $\Vcs$ is determined
to be \vcsresult. The differential cross section as a function
of the W$^-$ production angle is measured for the $\Sqqev$ and
$\Sqqmv$ final states. The results described in this paper
are consistent with the expectations from the Standard Model.

\bigskip\bigskip\bigskip\bigskip

\begin{center}
{\large {This paper is dedicated to the memory of Ben Shen}} 
\bigskip\bigskip \\

    {\large(To be submitted to Eur. Phys. J. C)}
\end{center}
\end{titlepage}

\begin{center}{\Large        The OPAL Collaboration
}\end{center}\bigskip
\begin{center}{
G.\thinspace Abbiendi$^{  2}$,
C.\thinspace Ainsley$^{  5}$,
P.F.\thinspace {\AA}kesson$^{  7}$,
G.\thinspace Alexander$^{ 21}$,
G.\thinspace Anagnostou$^{  1}$,
K.J.\thinspace Anderson$^{  8}$,
S.\thinspace Asai$^{ 22}$,
D.\thinspace Axen$^{ 26}$,
I.\thinspace Bailey$^{ 25}$,
E.\thinspace Barberio$^{  7,   p}$
% ...
}\end{center}\bigskip
\bigskip
$^{  1}$School of Physics and Astronomy, University of Birmingham,
Birmingham B15 2TT, UK
\newline
$^{  2}$Dipartimento di Fisica dell' Universit\`a di Bologna and INFN,
I-40126 Bologna, Italy
\newline
% ...
$^{  5}$Cavendish Laboratory, Cambridge CB3 0HE, UK
\newline
% ...
$^{  7}$CERN, European Organisation for Nuclear Research,
CH-1211 Geneva 23, Switzerland
\newline
$^{  8}$Enrico Fermi Institute and Department of Physics,
University of Chicago, Chicago IL 60637, USA
\newline
% ...
$^{ 21}$Department of Physics and Astronomy, Tel Aviv University,
Tel Aviv 69978, Israel
\newline
$^{ 22}$International Centre for Elementary Particle Physics and
Department of Physics, University of Tokyo, Tokyo 113-0033, and
Kobe University, Kobe 657-8501, Japan
\newline
% ...
$^{ 25}$University of Victoria, Department of Physics, P O Box 3055,
Victoria BC V8W 3P6, Canada
\newline
$^{ 26}$University of British Columbia, Department of Physics,
Vancouver BC V6T 1Z1, Canada
\newline
% ...
$^{  p}$ now at The University of Melbourne, Victoria, Australia
\newline
% ...

\section{Introduction}

From $1996-2000$ the LEP $\epem$ collider at CERN operated at centre-of-mass 
energies, $\roots$, above the threshold for $\WW$ production. This paper 
describes the OPAL measurements of the $\WW$ production cross section and W 
branching fractions using this data sample that corresponds to
a total integrated luminosity of $\totlumi$. The OPAL analysis of 
$\WW$ production and decay using data recorded at 
$\roots>190$\,GeV has not been published previously. 
For this paper the data recorded at 183\,GeV and above have been analysed using the final OPAL
detector calibration and W pair event selections. The results presented here
supersede the previous OPAL analysis of the data recorded at 
$\roots=183$\,GeV\cite{bib:xs183} and $\roots=189$\,GeV\cite{bib:xs189}.
The data collected close to the W pair production threshold 
($\roots=161$\,GeV and $172$\,GeV) have not been reanalysed and the 
corresponding results are described in \cite{bib:xs161,bib:xs172}. 
Furthermore, for the reasons explained in Section~\ref{sec:sellvlv}, 
the 183\,GeV $\WWlnln$ data
have not been reanalysed and the corresponding results are given
in \cite{bib:xs183}.

In this paper, $\WW$ production is defined in 
terms of the \CC\ class\cite{bib:yellow} of 
production diagrams. These diagrams, which correspond to $t$-channel $\nue$ exchange 
and $s$-channel $\Zz/\gamma$ exchange, provide a natural definition of 
resonant W-pair production. The contributions to the event rate
from non-\CC\ diagrams which lead to the same final states
as W-pair production (including interference with the
\CC\ set of diagrams) are treated as additive background.
In the Standard Model (SM), $\WW$ events are expected to decay into 
fully leptonic ($\lnln$), semi-leptonic ($\qqln$), or fully hadronic
($\qqqq$) final states with predicted SM branching fractions of 
10.6\,\%, 43.9\,\% and 45.6\,\% respectively\cite{bib:yellow}.
Here $\qq$ denotes a quark and an anti-quark and $\ell\nu$ denotes
a lepton/anti-lepton ($\ell$ = e, $\mu$, $\tau$) and an anti-neutrino/neutrino.
Three separate event selections, described in Section \ref{sec:selection},
are used to identify candidate $\WW$ events by their final state
topologies with $\lnln$ and $\qqln$ candidates classified according to
the charged lepton type. From the observed event rates in these ten channels
(6 $\lnln$, 3 $\qqln$ and $\qqqq$)
measurements of the W boson branching fractions and total $\WW$ production
cross section are obtained. The measured branching fraction to hadrons is used
to provide a determination of the CKM matrix element $\Vcs$. For the $\qqen$ 
and $\qqmn$ decay channels the charge of the W bosons can be identified from the
charge of the observed lepton. These events are used to determine the differential
cross section in terms of the $\mathrm{W}^-$ polar angle.

\section{Detector, Data and Monte Carlo}

\subsection{The OPAL Detector}

The inner part of the \Opal\ detector consisted of a 3.7 m diameter tracking volume within
a 0.435~T axial magnetic field. The tracking detectors included a silicon
micro-vertex detector, a high precision gas vertex detector and a large
volume gas jet chamber. The tracking acceptance corresponds to approximately
$|\cos\theta|<0.95$ (for the track quality cuts used in this study), 
where $\theta$ is the polar angle with respect to the $\mathrm{e}^-$ beam direction.
The transverse momentum resolution for muon tracks is 
approximately 
$\sigma_{p_{\mathrm T}}/p_{\mathrm T}=\sqrt{(0.02)^2+(0.0015p_{\mathrm T})^2}$ 
with $p_{\mathrm T}$ measured in GeV.
Lying outside the solenoid, 
the electromagnetic calorimeter (ECAL) 
consisting of $11\,704$ lead glass blocks
 had full acceptance in the range $|\cos\theta|<0.98$ and a relative
energy resolution for electrons of approximately 
$\sigma_E/E \approx 0.18/\sqrt{E}$ with $E$ measured in GeV.
The magnet return yoke was instrumented with streamer tubes which
served as the hadronic calorimeter. Muon chambers outside the 
hadronic calorimeter provided muon identification in the range
$|\cos\theta|<0.98$. 
Hermeticity for polar angles down to approximately 24\,mrad 
was achieved with forward detectors designed for measuring 
electrons and photons. 
Additional forward scintillator tiles were installed
in 1998 in order to extend the coverage for detection of minimum ionising 
particles\cite{bib:mipplug}. These forward scintillator tiles were used to
improve the $\lnln$ analysis for the $\roots \ge 189\,\GeV$ data samples. 
A detailed description of the \Opal\ 
detector can be found in \cite{bib:detector}.  

\subsection{Data Sample}

From 1996 onwards the centre-of-mass energy of the \LEP\ collider
was increased from 161~GeV to 209~GeV in several steps. 
The total integrated luminosity of 
the data sample considered in this paper, evaluated using small angle 
Bhabha scattering events observed in the silicon tungsten forward 
calorimeter~\cite{bib:lumi}, is $\totlumiwitherror$.
For the purpose of measuring the $\WW$ cross section these data
are divided into ten  $\roots$ ranges listed in 
Table~\ref{tab:lumi}. These ranges reflect the main energy steps as
the centre-of-mass energy was increased during \LEP\ operation above
the $\WW$ production threshold. 
  
 \begin{table}[htbp]
   \centering
    \begin{tabular}{c|c|c} \hline\hline
  Range/GeV  & $\langle\roots\rangle$/GeV  & $\cal{L}$/pb$^{-1}$ \\ \hline
160.0$-$165.0&161.30&  9.89\\
165.0$-$180.0&172.11& 10.36\\
180.0$-$185.0&182.68& 57.38\\
185.0$-$190.0&188.63&183.04\\
190.0$-$194.0&191.61& 29.33\\
194.0$-$198.0&195.54& 76.41\\
198.0$-$201.0&199.54& 76.58\\
201.0$-$202.5&201.65& 37.68\\
202.5$-$205.5&204.88& 81.91\\
205.5$-$209.0&206.56&138.54\\ \hline
Total & $-$ & 701.12\\
 \hline\hline
 \end{tabular}
 \caption{The energy binning used for the \WW\ cross section measurements.
  The $\roots$ range covered by each bin, the mean luminosity-weighted 
  value of $\roots$ and the corresponding integrated luminosity, 
  ${\cal{L}}$, are listed.
 \label{tab:lumi} }
 \end{table}

\subsection{Monte Carlo}

A number of Monte Carlo (MC) samples, all including a 
full simulation\cite{bib:GOPAL} of the \Opal\ detector, 
are used to model the signal and background processes.
For this paper the main MC samples for four-fermion final states
consistent with coming from the process 
$\epem\rightarrow\WW$ are generated using the 
\KANDY~\cite{bib:KandY} program. 
\KANDY\ includes exact $\cal{O}(\alpha)$ YFS 
exponentiation\cite{bib:YFS} for the \WW\ production process, 
with $\cal{O}(\alpha)$ electroweak non-leading (NL) corrections combined 
with YFS exponentiated ${\cal{O}}(\alpha^3)$ leading logarithm (LL) 
initial state radiation (ISR). Final state radiation (FSR) from leptons 
is implemented in \PHOTOS~\cite{bib:Photos}  and radiation from
the quark induced parton-shower is performed by \JETSET~\cite{bib:Jetset}. 
The hadronisation within the \JETSET\ model is tuned to OPAL data 
recorded at the $\Zz$ resonance\cite{bib:qqqqQCD}. For the studies of 
systematic uncertainties the \JETSET\ hadronisation
model is compared with the predictions from \HERWIG~\cite{bib:herwig} and 
\ARIADNE~\cite{bib:ariadne}.

The \KANDY\ generator is also used
to produce event weights such that generated events can be reweighted
to correspond to the \CC\ set of diagrams alone. The difference
between the full set of four-fermion diagrams 
and the \CC\ diagrams alone is used to obtain the 
four-fermion background which includes the effects of interference with
the \CC\ diagrams. 

The \KORALW\ program\cite{bib:KoralW} is used to simulate the background
from four-fermion final states which are incompatible with
coming from the decays of two W-bosons (e.g. $\epem\rightarrow\qq\mpmm$). 
The two-fermion background processes 
$\epem\rightarrow\Zz/\gamma\rightarrow\mpmm$,
$\epem\rightarrow\Zz/\gamma\rightarrow\tptm$
and $\epem\rightarrow\Zz/\gamma\rightarrow\qq$ are
simulated using \KK~\cite{bib:kk}. The two fermion process
$\epem\rightarrow\Zz/\gamma\rightarrow\epem$ is simulated using 
\BHWIDE~\cite{bib:BHWIDE}.
Backgrounds from two-photon interactions are evaluated using 
\PYTHIA~\cite{bib:pythia}, 
\HERWIG, \PHOJET~\cite{bib:PHOJET}, \BDK~\cite{bib:BDK} 
and the Vermaseren program~\cite{bib:VERMASEREN}.

The SM predictions for the \CC\ $\epem\rightarrow\WW$ cross sections 
above the $\WW$ threshold region are
obtained from the $\YFSWW$\cite{bib:YFSWW} 
and the $\RACWW$\cite{bib:RacoonWW} programs. \RACWW\
is a complete $\cal{O}(\alpha)$ $\epem\rightarrow 4f\gamma$ calculation in the double
pole approximation with ISR treated using a structure function approach.
The \YFSWW\ program provides the $\WW$ calculations in \KANDY.
\YFSWW\ and \RACWW\ yield nearly identical predictions for the $\WW$ cross sections with an estimated theoretical uncertainty 
of approximately 0.5\,\%\cite{bib:lep4f}. For W-pair production
near threshold (the 161\,GeV and 172\,GeV data) the leading- and double-pole 
approximations used in \YFSWW\ and \RACWW\ respectively are no longer valid and
the predictions are obtained from both calculations using the Improved Born 
Approximation where the theoretical uncertainty is approximately 2\,\%.

\section{\boldmath $\epem\rightarrow\WW$ Event Selection}

\label{sec:selection}

The selection of $\WW$ events proceeds in three stages,
corresponding to the three $\WW$ decay topologies:
$\WWlnln$, $\WWqqln$ and $\WWqqqq$. The selections are mutually
exclusive with only events failing the $\WWlnln$ selection being
considered in the  $\WWqqln$ selection, and only events which 
are not selected as $\lnln$ or $\qqln$ being considered for the
$\WWqqqq$ selection. The event selections are essentially unchanged 
from those described in detail in\cite{bib:xs189} (and references therein) 
although the $\WWlnln$ selection now incorporates features used in the \OPAL\
analysis of di-lepton events with significant 
missing transverse momentum\cite{bib:acopdilepton}.

In the centre-of-mass energy range $\roots = 161-209$ GeV, the 
luminosity-weighted average \CC\ W-pair selection efficiencies for the 
$\lvlv$, $\qqlv$ and $\qqqq$ decay channels are 84\,\%, 84\,\% and 
86\,\% respectively. This corresponds to a total efficiency of 85\,\%.  
The selection efficiencies, broken down into the different lepton flavours
are summarised in Table~\ref{tab:lumi_weighted_effic}. For 
the data samples away from the W-pair threshold the selection 
efficiencies depend only weakly on centre-of-mass energy.
The main features of the selections and associated systematic uncertainties
are described below in Sections \ref{sec:sellvlv}$-$\ref{sec:selqqqq}.

 \begin{table}[htbp]
   \centering
    \begin{tabular}{c|cccccc|ccc|c} \hline\hline
 Event & \multicolumn{10}{c}{Efficiencies[\%] for $\WW\rightarrow$} \\
 Selection & \Gevev & \Gmvmv & \Gtvtv & \Gevmv & \Gevtv & \Gmvtv & \Gqqev & \Gqqmv & \Gqqtv & \Gqqqq \\ \hline\hline
\Sevev& 74.1&  0.0&  0.8&  0.4&  6.6&  0.1&  0.0&  0.0&  0.0&  0.0\\
\Smvmv&  0.0& 77.9&  0.7&  1.4&  0.1&  6.7&  0.0&  0.0&  0.0&  0.0\\
\Stvtv&  0.7&  0.7& 48.1&  0.7&  4.9&  5.6&  0.0&  0.0&  0.0&  0.0\\
\Sevmv&  2.6&  0.4&  1.4& 76.5&  6.2&  6.9&  0.0&  0.0&  0.0&  0.0\\
\Sevtv& 10.3&  0.0& 11.5&  5.6& 64.2&  1.2&  0.0&  0.0&  0.0&  0.0\\
\Smvtv&  0.2&  9.5&  8.4&  4.3&  0.8& 61.5&  0.0&  0.0&  0.0&  0.0\\
 \hline
\Sqqev&  0.0&  0.0&  0.0&  0.0&  0.2&  0.0& 84.3&  0.1&  4.0&  0.0\\
\Sqqmv&  0.0&  0.0&  0.0&  0.0&  0.0&  0.1&  0.2& 88.3&  4.4&  0.1\\
\Sqqtv&  0.0&  0.0&  0.2&  0.0&  0.0&  0.0&  4.3&  4.4& 61.5&  0.5\\
 \hline
\Sqqqq&  0.0&  0.0&  0.0&  0.0&  0.0&  0.0&  0.0&  0.1&  0.8& 85.9\\
 \hline\hline
 \end{tabular}
 \caption{\label{tab:lumi_weighted_effic} The luminosity-weighted average
 selection efficiencies for the CC03 processes for $\roots=161-209$\,\GeV.
 The efficiencies include corrections for detector occupancy and tracking
 inefficiencies as described in the text.}
 \end{table}

\subsection{\boldmath Selection of $\WWlnln$ events}
\label{sec:sellvlv}

The $\WWlnln$ process results in an event with two charged leptons, 
not necessarily of the same flavour, and significant missing 
momentum. This characteristic event topology is 
of interest both for measuring aspects of W physics and for exploring the 
potential production of new particles leading to the same experimental 
signature. The $\WWlnln$ event selection described here first requires 
events to be selected by the general event selection used by OPAL to
search for new particles such as pair production of super-symmetric 
particles which decay leptonically\cite{bib:acopdilepton}. This selection
identifies events consistent with there being two charged leptons and 
significant missing transverse momentum. From this 
sample cuts are applied to identify events consistent with being from the 
$\WWlnln$ process. This event selection takes advantage of changes to the
OPAL detector made in 1998. Consequently the data from centre-of-mass 
energies of 161\cite{bib:xs161}, 172\cite{bib:xs172} and 
183\,GeV\cite{bib:xs183} have not been reanalysed.

The general $\llnunu$ event selection is described in detail in 
\cite{bib:acopdilepton} and references therein. 
The selection is formed by requiring that an event 
be selected by either of two independent event selections, 
referred to in ~\cite{bib:acopdilepton} as Selection I and Selection II.  
Both event selections require evidence for significant missing 
transverse momentum and are designed to minimise background 
contributions from SM processes which can lead to an experimental 
signature of two charged leptons and significant missing transverse momentum. 
In the case of background processes, significant missing transverse momentum
can arise from a number of sources: secondary neutrinos in tau decays; 
mis-measurement of the lepton energies and directions; or where 
high transverse momentum particles are incident on poorly instrumented 
regions of the detector.

Selection I is designed to retain efficiency for events with low visible 
energy. Selection II is designed for measuring 
$\WWlnln$ events which usually have substantial visible energy; the 
selection criteria have been optimised to maximise the 
statistical power (efficiency multiplied by purity) treating 
\CC\ $\WWlnln$ as signal and SM processes other 
than $\lnln$ as background. 
For both Selection I and Selection II particular care is taken to reject 
events with fake missing momentum due to detector effects.
Neither selection attempts  
to reduce the sensitivity to non-\CC\ sources of $\lnln$ events 
with two detected leptons. There is a large overlap in the 
expected acceptance of the two selections: from the selected MC event sample, 
6\,\% of events are selected exclusively by Selection I and 6\,\% exclusively
by Selection II. Conversely, of the MC SM background events from 
processes other than $\lnln$, 9\,\% pass both selections,
32\,\% exclusively pass Selection I and 59\,\% exclusively pass Selection II.

Both selections are cut-based and rather involved\cite{bib:acopdilepton}, 
and only an outline of the main points is given here. The most significant 
variables used are: $x_{min}$ ($x_{max}$), the momentum of the 
lower (higher) momentum charged lepton candidate scaled 
to the beam energy;
$\xT$, the magnitude of the missing momentum scaled to the beam 
energy; $\acop$, the supplement of the azimuthal opening angle;
$\theta_p^{\mathrm{miss}}$, the polar angle of the missing
momentum vector; 
$\pz$, the magnitude of the $z$ component of the missing 
momentum;
$\ptaxic$, the component of the missing transverse momentum 
that is perpendicular to the event thrust axis in the transverse plane;
and $\athet = \tan^{-1}[\ptaxic /\pz]$.

Selection I is based on three main requirements:
\begin{itemize}
\item{evidence that a pair of charged leptons is produced, where
at least one must have  $\pT$ exceeding 1.5 GeV and must satisfy 
requirements on lepton identification and isolation;}
\item{evidence of statistically significant missing transverse momentum.
For large acoplanarity events, $\acop > \pi/2$, $\xT$ is required to 
exceed 0.045. For $\acop < \pi/2$, {i.e.} events where the leptons 
are more back-to-back, a combination of cuts on $\xT$, $\aT$ and 
$\athet$ is used. The cuts depend on the di-lepton 
identification information;}
\item{a veto on events with fake missing transverse momentum using the 
      detectors in the forward region of the detector.}
\end{itemize}

Selection I is designed as a general selection for di-lepton events with
missing transverse momentum. In order to isolate events consistent with 
the process $\WWlnln$, additional cuts are applied in this analysis 
to remove events 
which have relatively low 
missing transverse momentum 
(an important region for SUSY and other new particle searches but not
for W-pair production):
\begin{itemize}
  \item{events are rejected if $x_{max} < 0.1$;}
  \item{if $\xT < 0.2$, $|\cos\theta_p^{\mathrm{miss}}| > 0.7$ and 
         $x_{min} < 0.3$, events are rejected if either $x_{max} < 0.15$ or 
         $\acop < \pi/2$ and $\athet<0.1$;}
  \item{for events with only one reconstructed isolated charged lepton 
        candidate, events are rejected if the net momentum of the additional 
        tracks and clusters not associated to the lepton divided by their 
        invariant mass is less than 4.}
\end{itemize}

Selection II starts from a preselected sample of low multiplicity 
events and makes little use of lepton identification 
information in the event selection procedure. The first stage of the selection is 
to apply a cone jet-finding algorithm\cite{bib:cone} using a 
cone half-opening angle of $20^\circ$ and a jet energy threshold of 2.5 GeV. 
The majority (90\,\%) of $\WWlnln$ events are reconstructed in the di-jet category. 
For events reconstructed as two jet events, the three most important selection 
criteria are: 
\begin{itemize}
  \item{evidence for missing transverse momentum defined by 
     requiring that $\xT$ should exceed 0.05 by a 
     statistically significant margin;} 
  \item{for low acoplanarity events $\aT$ should exceed 0.020, primarily to 
     reject events where the missing momentum arises from 
     secondary neutrinos from tau decays;} 
  \item{a veto on activity in the forward region similar to Selection I}.
\end{itemize}
Additional selections targeted at three-jet events (often $\WWlnln\gamma$) 
and single jet events (one observed lepton plus evidence for the presence 
of another lepton) are used to improve the overall selection efficiency.

Events are classified as one of the six possible di-lepton types. 
For events selected by Selection II, the event classification uses 
both particle identification information and kinematic 
information as described in reference~\cite{bib:xs189}.
For events selected exclusively by Selection I the di-lepton classification is
based on the lepton identification information only. 

\subsubsection{\boldmath $\WWlnln$ Selection Systematic Uncertainties}

\par
\noindent
{\bf Efficiency Uncertainties:}
The OPAL trigger and pretrigger systems 
provide a highly redundant
and efficient trigger for $\WWlnln$; studies indicate
that the trigger inefficiency for events selected by these event
selections is negligible.
The $\WWlnln$ event selection efficiencies are limited mainly by the
geometrical acceptance of the detector and the defined kinematic 
acceptance. The latter is implicit in the requirement that the observed final 
state particles have a net visible transverse momentum 
which significantly exceeds that which could be explained by undetected 
particles at low polar angles. The 
detector acceptance is well understood and factors affecting the kinematic acceptance 
such as momentum and energy scales and resolutions are adequately modelled 
by the MC simulation. 
Extensive studies have been carried out
comparing distributions of the event selection
variables in data with MC.
In general, reasonable agreement is found and
quantitative estimates of the individual systematic effects are
small compared to the statistical errors.
In particular, the critical distributions
associated with requiring missing transverse momentum,
such as the $\aT$ and the $\xT$ distributions are well modelled. 
As an example, the single most important
cut in the two ``jet'' part of Selection II is the cut on $\aT$
which leads to a relative loss in the $\WWlnln$ efficiency of 1.1\%.
A conservative estimate of the systematic error on
the $\aT$ scale of 1\% leads to a systematic uncertainty of 0.04\% 
on the overall efficiency. As a result of such studies, an overall global event 
selection efficiency systematic uncertainty corresponding to 5\,\% of the 
inefficiency prior to occupancy corrections is assessed. 
This systematic uncertainty is taken to be fully correlated 
among centre-of-mass energies 
and ranges from 0.7\,\% at 189 GeV to 0.8\,\% at 207 GeV.

\par
\noindent
{\bf Detector Occupancy:}
The $\WWlnln$ event selection is sensitive to hits in the various sub-detectors
which do not arise from the primary $\epem$ interaction, termed ``detector occupancy''. 
Backgrounds from the accelerator, cosmic-ray muons, or 
electronic noise can lead to additional hits, energy deposition and
even reconstructed tracks being superimposed on triggered data events. 
These detector occupancy effects are simulated by adding to the reconstructed
MC events the hits, energy depositions and additional ``jets'' found in randomly 
triggered\cite{bib:opal_trigger} beam-crossing data events corresponding to the same 
centre-of-mass energy. 
The detector occupancy corrections are included in the quoted efficiencies of
Table~\ref{tab:lumi_weighted_effic}. They reduce the overall efficiency and range from 
$-0.4\,\%$ at 189 GeV to $-1.0\,\%$ at 207 GeV. The variation is due to higher 
beam-related backgrounds at the highest energies. In order to take into account 
residual deficiencies in the implementation of these post event reconstruction
corrections, a systematic uncertainty 
amounting to one half of the correction is assigned.


The overall $\lvlv$ efficiency systematic uncertainties (for all final states combined) 
range from 0.8\,\% to 1.0\,\% for centre-of-mass energies of $189-209$\,GeV.

\bigskip
\noindent
{\bf Background Uncertainties:}
%
%
There are three main sources of background in the $\WWlnln$ selection:
\begin{itemize}
\item{ {\bf\boldmath Non-$\llnunu$ Background:} Events from processes with no 
  primary neutrinos which manage to fake the missing transverse 
  momentum signature. Important sub-components are di-lepton production, in particular tau-pairs, 
  multi-peripheral two-photon processes and the four-fermion $\eeff$ processes.}
\item{ {\bf Non-interfering four-fermion background:} $\llnunu$ final states arising from processes 
such as $\ZzZz$ with primary neutrinos in the final state and with lepton and neutrino 
flavours incompatible with WW production (e.g. $\mu^+ \mu^- \nu_\tau \overline{\nu}_\tau$).}
\item{ {\bf Interfering four-fermion background:} The $\llnunu$ final states 
    relevant to $\WWlnln$ also have significant contributions from diagrams beyond those of 
    \CC\ W-pair production, such as $\Wenu$, $\Zee$, $\ZzZz$ and 
$\Zz\nue\nuebar$. These contributions, 
    which can 
    also interfere with the \CC\ diagrams, are treated as an additive 
     background.}
\end{itemize}
For the centre-of-mass energy range $\roots=161-209$\,GeV, the 
luminosity-weighted average expected background cross sections are listed 
in Table~\ref{tab:backgrounds}.

 \begin{table}[htbp]
   \centering
    \begin{tabular}{c|cccccc|ccc|c} \hline\hline
 Source of & \multicolumn{10}{c}{Background [fb] in selection} \\
 Background & \Sevev & \Smvmv & \Stvtv & \Sevmv & \Sevtv & \Smvtv & \Sqqev & \Sqqmv & \Sqqtv & \Sqqqq \\ \hline\hline
\Blvlv&  20.&  17.&  18.&  21.&  31.&  17.&   0.&   0.&   0.&   0.\\
\Bqqlv&   0.&   0.&   0.&   0.&   0.&   0.&  61.&   3.&  73.&   0.\\
\Bqqqq&   0.&   0.&   0.&   0.&   0.&   0.&   0.&   1.&   6.& 493.\\ \hline
\Bllll&   1.&   1.&   5.&   0.&   3.&   2.&   1.&   0.&   1.&   0.\\
\Bqqll&   0.&   0.&   0.&   0.&   0.&   0.&  38.&  30.&  77.&  49.\\
\Bqqvv&   0.&   0.&   0.&   0.&   0.&   0.&   1.&   1.&  36.&   0.\\
\Bll  &   2.&   2.&   5.&   1.&   5.&   3.&   2.&   1.&   5.&   0.\\
\Bqq  &   0.&   0.&   0.&   0.&   0.&   0.&  41.&  23.&  78.&1340.\\
\BeeX &   0.&   0.&   7.&   0.&   2.&   1.&   7.&   2.&   3.&   0.\\
 \hline\hline
Total&  23.&  21.&  35.&  23.&  41.&  23.& 152.&  63.& 280.&1882.\\
error&   2.&   3.&   4.&   2.&   3.&   3.&  10.&   5.&  32.& 100.\\
 \hline\hline
 \end{tabular}
   \caption{Luminosity-weighted average background cross sections [fb]
    in the different event selection categories. The background 
    cross sections for the $\qqtn$ selection include the corrections
    described in the text. The quoted errors include both statistical
    and systematic uncertainties. \label{tab:backgrounds}}
 \end{table}

The overall systematic uncertainties on the background cross sections for 
each di-lepton class and at each centre-of-mass energy are calculated by 
summing up the contributions in the following categories. The uncertainties within each 
category are assumed to be fully correlated among di-lepton channels and 
centre-of-mass energies.
\begin{itemize}
\item{For events from di-lepton production the theoretical uncertainties are negligible.
      In this case it is simulation of the detector response that dominates the 
      uncertainty on the background. Events are selected due to either
      mis-measurements of the variables used in the selection or from the tails of 
      the $\tptm$ decay distributions. An overall background systematic uncertainty of 10\,\% is 
      assessed.}
\item{A 5\,\% systematic uncertainty is assigned to the background expectations
      from genuine $\llnunu$ events coming both from non-interfering four-fermion 
      background final states and from the non-\CC\ contribution to final states where the four
      fermions are compatible with being from W-pair production.}
\item{A 10\,\% systematic uncertainty is assigned to the background expectations from $\eeff$ 
       and the remaining small contributions from other four-fermion processes,
       reflecting the theoretical error on simulation of processes like $\Zee$.}
\item{For events from the multi-peripheral $\eeX$ process an uncertainty of 30\,\% is assigned. 
      The uncertainty reflects the size of the discrepancy in the
      modelled number of events exclusively rejected using the forward
      scintillating tiles, a category of events dominated by multi-peripheral
      backgrounds.} 
\end{itemize}

\par
\noindent
{\bf Event Classification Uncertainties:}
There are two aspects to the di-lepton flavour classification of selected 
$\WWlnln$ candidates. 
Firstly, the algorithms for leptons to be identified  
as electrons, muons or hadronically decaying taus. These make use of many of the techniques of 
lepton identification used by OPAL in studies at the $\Zz$.
Secondly, the kinematic re-classification algorithm based on scaled momentum which re-classifies 
soft leptons identified as electrons or muons as probable secondary leptons from taus, 
and uses electromagnetic calorimeter and muon information to re-assess whether 
highly energetic leptons initially not identified as electrons or muons are 
more consistent kinematically with prompt electrons or muons.
The classification efficiency 
systematic uncertainty for genuine electrons and muons 
is assessed to be 2\,\% based on the understanding of the lepton identification
information in the large $\epem\rightarrow\lplm$ samples recorded at \LEPI. 
The kinematic re-classification, which relies  mainly on measurement of the lepton energy,
reduces the systematic uncertainties on the efficiencies for the individual
final state lepton channels to the 1\,\% level. 
In the extraction of the SM parameters that follows it has been verified that the 
effects of the $\lnln$ classification systematic uncertainties are small. 
Nevertheless, the effects of the classification systematic uncertainties 
and correlations are included in the analysis.

\subsubsection{\boldmath $\WWlnln$ Results}

Using the \KANDY\ MC samples the
luminosity-weighted average \CC\ $\WWlnln$ event selection efficiency 
in the 189$-$209\,GeV centre-of-mass energy range is estimated to be $(84.7 \pm 0.8)\,\%$.
The inclusive selection efficiencies for the different centre-of-mass 
energies are listed in Table~\ref{tab:lvlv_xs}. 
The efficiencies for the different final states 
depend mostly on the number of taus present. The luminosity-weighted average efficiencies are
89.4\,\%, 83.2\,\% and 71.9\,\% for final states with zero, one and two taus respectively.
For the 189$-$209\,GeV data the selection efficiency does not depend strongly
on centre-of-mass energy. 
The luminosity weighted efficiencies of 
the \WWlnln\ 
selection for the individual channels are given in 
Table~\ref{tab:lumi_weighted_effic}. 
The efficiencies/numbers of expected events
in all tables include the detector occupancy corrections 
described above.

\begin{figure}
 \begin{center}
  \epsfig{file=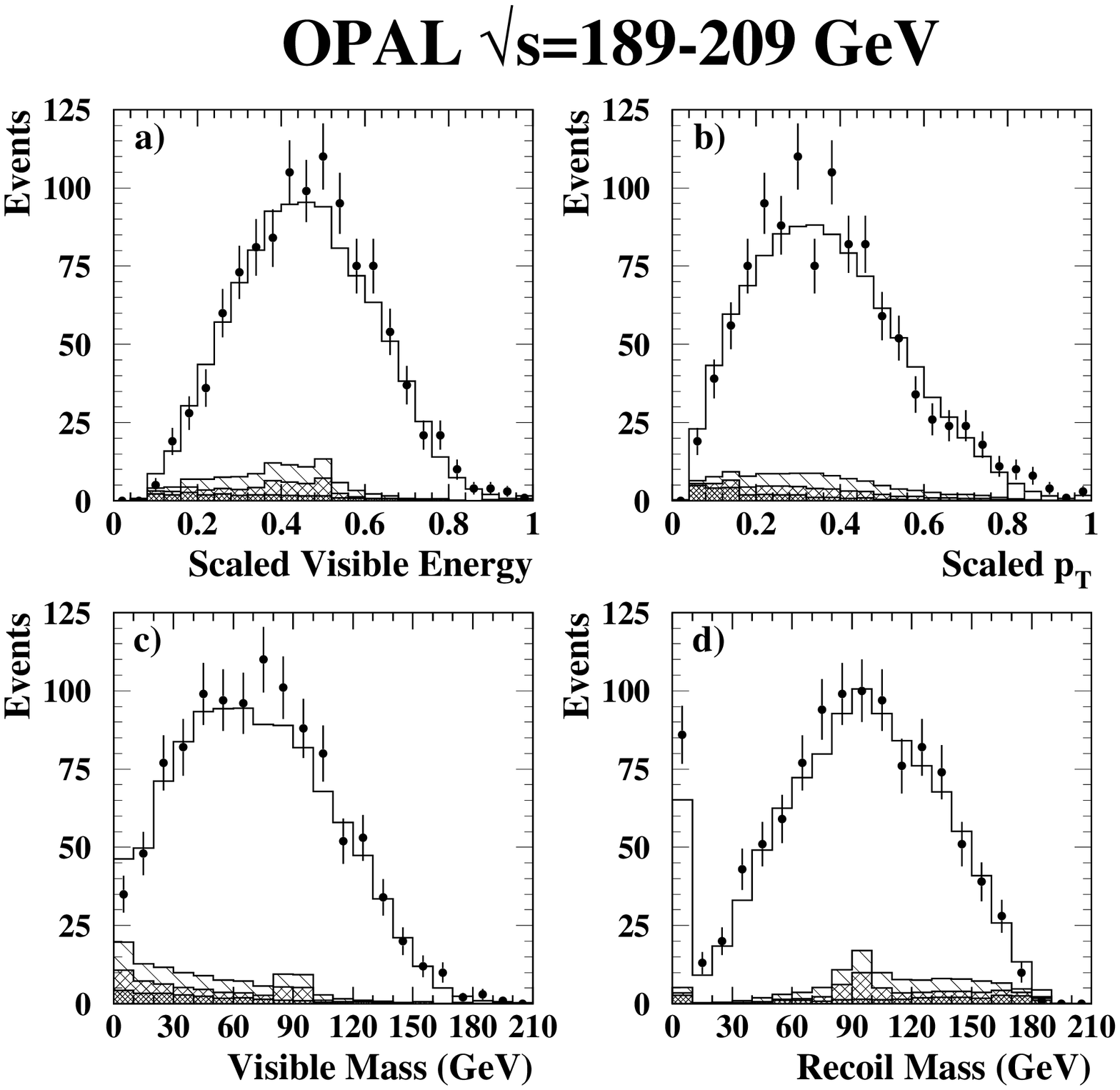,width=\textwidth}
 \caption{ Distributions of (a) the total visible energy in the event scaled
to the centre-of-mass energy, (b) the magnitude of the net visible transverse momentum 
in the event scaled to the beam energy, (c) the reconstructed total visible
invariant mass of the event, and (d) the invariant mass of the system recoiling
against the visible system. All plots show the selected $\lnln$ events
for the combined sample from
data recorded at $\sqrt{s}=189-209$\,GeV. In (d) the events in the first
bin are where the reconstructed recoil mass squared is negative.
The data are shown as the points with error bars (statistical errors only).
The total Standard Model MC prediction is shown by the unshaded histogram.
The background components are also shown: interfering $\lnln$ (singly-hatched),
non-interfering $\llnunu$ (cross-hatched) and two fermion/multi-peripheral
(densely cross-hatched). The MC is normalised to the 
integrated luminosity of the data. 
\label{fig:lnlnsel}
     }
 \end{center}
\end{figure}

In total, 1188 events are selected as \WWlnln\ candidates 
compared to the SM expectation of $1138\pm9$ 
(the numbers refer to the entire data set from $161-209$\,GeV). 
Figure~\ref{fig:lnlnsel} shows kinematic distributions for
reconstructed $\WWlnln$ event samples. The data distributions are in good 
agreement with the MC expectations. 
The numbers of selected $\lnln$ events at each energy are used to
determine the cross sections for $\epem\rightarrow\WW\rightarrow\lnln$
given in Table~\ref{tab:lvlv_xs}. 
The measured cross sections are in agreement with the SM expectations.

 \begin{table}[htbp]
   \centering
    \begin{tabular}{c|cccc|c|c} \hline\hline
  $\roots$ & $\cal{L}$ & $N$ & Efficiency & Background     & $\sigma(\WW\rightarrow\lnln)$ & SM \\
   $[\mathrm{GeV}]$    & [pb$^{-1}$]  & [events]  & [\%]   & [events] & [pb] & [pb]\\ \hline\hline
161.30&  9.9&    2& 65.4$\pm$  2.0&  0.2$\pm$  0.0& 0.28$\pm$ 0.22$\pm$ 0.01& 0.38\\
172.11& 10.4&    8& 78.2$\pm$  2.6&  0.8$\pm$  0.3& 0.89$\pm$ 0.35$\pm$ 0.03& 1.28\\
182.68& 57.4&   78& 78.1$\pm$  2.3&  4.9$\pm$  1.5& 1.63$\pm$ 0.20$\pm$ 0.05& 1.62\\ \hline
188.63&183.0&  295& 86.1$\pm$  0.8& 28.1$\pm$  0.7& 1.69$\pm$ 0.11$\pm$ 0.02& 1.72\\
191.61& 29.3&   56& 85.3$\pm$  0.8&  4.9$\pm$  0.2& 2.04$\pm$ 0.30$\pm$ 0.02& 1.75\\
195.54& 76.4&  145& 85.1$\pm$  0.8& 13.0$\pm$  0.4& 2.03$\pm$ 0.19$\pm$ 0.02& 1.78\\
199.54& 76.6&  138& 84.8$\pm$  0.8& 13.6$\pm$  0.4& 1.91$\pm$ 0.18$\pm$ 0.02& 1.79\\
201.65& 37.7&   86& 83.9$\pm$  0.9&  7.1$\pm$  0.2& 2.50$\pm$ 0.29$\pm$ 0.03& 1.80\\
204.88& 81.9&  141& 83.5$\pm$  1.0& 16.3$\pm$  0.5& 1.82$\pm$ 0.17$\pm$ 0.02& 1.81\\
206.56&138.5&  239& 83.5$\pm$  1.0& 27.8$\pm$  0.8& 1.83$\pm$ 0.13$\pm$ 0.02& 1.81\\
 \hline\hline
 \end{tabular}
 \caption{\label{tab:lvlv_xs} Measured cross sections for
 the \CC\ process $\epem\rightarrow\WW\rightarrow\lvlv$.
 For the $\lvlv$ selection the data below $\roots=188.63$\,GeV have not been reanalysed and the 
 results are taken from~\cite{bib:xs172,bib:xs183,bib:xs161}. 
 The errors on the cross sections are statistical
 and systematic.
 The numbers of selected events,
 the $\lvlv$ selection efficiencies and the expected numbers of
 background events are also listed.
 The backgrounds include a small contribution from semi-leptonic 
 $\WW$ decays which for the cross sections
 are taken to be fixed to their SM expectations.
 }
 \end{table}

\subsection{\boldmath Selection of $\WWqqln$ events}

\label{sec:selqqlv}

The \WWqqln\ selection consists of three separate selections, one
for each type of semi-leptonic decay. Only those events which are
not already selected as \lnln\ candidates are considered by these
selections. For each of the \WWqqen, \WWqqmn, and \WWqqtn\ event
selections, the main part is a relative likelihood method to reject the
potentially large \eetoqq\ background. In the first stage, the 
$\WWqqen$ and $\WWqqmn$ likelihood selections are performed. The 
\WWqqtn\ likelihood selection is only applied to those events which 
have not already been selected. Finally, events passing either the 
$\WWqqen$ or the $\WWqqmn$ selections may then be reclassified as \WWqqtn\ candidates. 

The \WWqqln\ event selections used here are almost 
identical to those described in previous \OPAL\ 
publications~\cite{bib:xs183,bib:xs189}.
However, using the entire 
\OPAL\ \WW\ data has resulted in an improved understanding of the
selection efficiencies and backgrounds. Using the improved 
estimates of the systematic uncertainties, the cut on the 
relative likelihood variable used to select $\qqtn$ 
candidates was re-optimised to minimise the 
total uncertainty (statistical and systematic) for this
channel.  As a result the cut on the likelihood was raised from
0.5 to 0.8 which reduces the efficiency by about 5\,\%. This loss in efficiency
is more than compensated by the factor of two reduction in background and 
the corresponding reduction in the associated systematic uncertainties.

\subsubsection{Event Selection}

The $\WWqqln$ event selection utilises the distinct topology of $\WWqqln$ 
events; missing energy and a high energy (usually isolated) lepton. 
The selection consists of six stages, 
which can be summarised as:
\begin{itemize}
   \item {\bf loose preselection:} a loose preselection to remove events 
         with low multiplicity or little visible energy.
   \item {\bf lepton candidate identification:} identification of the 
         observed track in the event which is most consistent with being from the 
         leptonic decay of a W boson. Candidate lepton tracks
         are identified for each of the $\qqen$, $\qqmn$ and $\qqtn$ 
         hypotheses.  
   \item {\bf preselection:} different sets of cuts are applied
         for \WWqqen, \WWqqmn, and \WWqqtn\ to remove events clearly
         incompatible with being signal (e.g. events are rejected if
         the total visible energy in the event is less than 0.3 of the
         centre-of-mass energy).
   \item {\bf relative likelihood selection:} different relative likelihood 
         selections are used to identify \WWqqen, \WWqqmn, and \WWqqtn\
         candidates.
         The probability density functions used in the likelihood selections 
         are obtained from MC at the different centre-of-mass energies. The
         variables used are either related to the properties of the lepton
         candidate (e.g. the lepton energy and degree 
         of isolation) or the kinematic properties of the event (e.g. the
         total visible energy and the magnitude of the missing momentum).
   \item {\bf decay classification:} 
         identification of \qqtn\ candidates from events which were 
         originally selected as \qqen\ or \qqmn.
   \item {\bf four-fermion background rejection:}
         rejection of four-fermion backgrounds $\qqll$, $\Wenu$, $\Zee$ and
         $\qqnn$. 
\end{itemize}
The first four stages, described in detail in \cite{bib:xs172}, 
are optimised for the rejection of the \eetoqq\ background which, 
for the centre-of-mass energies considered here,
has an expected cross section of between four and seven times larger than the
W-pair production cross section. The most important feature of the selection
is the looseness of the identification of possible lepton candidates. For both the \WWqqen\ and  
\WWqqmn\ selections the track which is most consistent with being from a 
leptonic W-decay is identified. The lepton track identification is based 
on an absolute likelihood taking into account momentum, isolation and 
lepton identification variables. To avoid associated systematic uncertainties 
only very loose cuts are placed on the lepton identification likelihood. 
The
lepton identification likelihood is then used as one of the input variables 
in the likelihood event selection. In this way the presence of 
{\em either} a good isolated lepton candidate or significant missing
transverse momentum is usually sufficient for an event to be selected.
This redundancy leads to high efficiency and reduces the dependence 
of the selection on the detailed simulation of the events and, consequently, 
leads to relatively small systematic uncertainties.  

Because of the limited use of lepton identification information, 
approximately $33\,\%$ of \WWqqtn\ events are accepted by at least one of the
$\qqen$ and $\qqmn$ likelihood selections. In addition, approximately 
$4\,\%$ of the \WWqqen\ and \WWqqmn\ events pass both the 
$\qqen$ and $\qqmn$ likelihood selections. Such events usually result from
there being a genuine electron from a W-boson decay and a track from one of the
jets being tagged as muon-like, or vice versa.
Consequently additional
likelihood selections, based primarily on lepton identification variables
and track momentum, are used to categorise events passing the 
$\qqen$ and $\qqmn$ likelihood selections into the three possible 
leptonic W-decay modes. The largest systematic uncertainties in the
efficiencies for selecting \WWqqln\ events are associated with this step.

Only events which failed the 
\WWqqen\ and \WWqqmn\ likelihood are passed to the 
\WWqqtn\ event selection.
The \WWqqtn\ event selection consists of separate selections for 
four possible tau decay signatures: $\tau\rightarrow\mathrm{e}\nu\nu$, 
$\tau\rightarrow\mu\nu\nu$, single prong hadronic decay modes and 
three prong hadronic decay modes. The main difference between these
selections is the power of the variables used to identify possible
tau decay products and the relative level of backgrounds.
An event is considered a $\qqtn$ candidate if it passes any one of these
four selections. 

Because the \WWqqln\ likelihood selections are designed to reject the
dominant \eetoqq\ background they have a significant efficiency 
for other four-fermion processes, {e.g.} \qqen\ final states
produced by the single W (\Wenu) diagrams and \qqll\ production
(mainly via $\eetoZZ$). Additional four-fermion background rejection  
cuts are applied to events passing the likelihood selections to reduce 
backgrounds from these processes.
The four-fermion background rejection consists of three separate parts. 
Cuts are applied to selected \qqen\ and \qqmn\ candidates to
reduce backgrounds from \qqee\ and \qqmm\ final states 
where both leptons are observed in the detector. 
Because of the lack of a clear signature for a lepton in \WWqqtn\
events, the selection places more weight on missing transverse 
energy to reject $\eetoqq$. Consequently the $\WWqqtn$ selection accepts 
approximately 40\,\% of hadronically decaying 
single W events ($\Wenu\rightarrow\qqen$). In these events the electron is usually produced in the
far forward region beyond the experimental acceptance and
a fragmentation 
track is mis-identified as a $\tau$ lepton decay product. To reduce 
this background, 
an additional likelihood selection is applied which separates 
\WWqqtn\ from \Wenu. This also rejects background from
$\epem\rightarrow\qqnn$. 
Background in the \WWqqen\ selection 
from the \Zee\ final state, where the \Zz\ decays hadronically and 
one electron is far forward, is reduced with two kinematic fits,
the first using the hypothesis that the event is \WWqqen\ and 
the second using the \Zee\ hypothesis.

In addition to the likelihood selections, 
cut based selections are used to identify \WWqqen\ and \WWqqmn\
events where the lepton track is either poorly reconstructed or 
is beyond the tracking acceptance. These `trackless' selections 
require clear evidence of an electron or muon
in the calorimeter or muon chambers consistent with the kinematics of a
\WWqqln\ event, without explicitly demanding a reconstructed track.
These additional selections improve the overall efficiency by approximately
3\,\% (5\,\%) for \WWqqen\ (\WWqqmn) events, and more importantly result
in a reduction in the systematic uncertainties associated with the modelling of 
the forward tracking acceptance.

\subsubsection{Systematic uncertainties}

\begin{table}[htbp]
\begin{center}
\begin{tabular}{ll|c|c|c|c} \hline\hline
 \multicolumn{2}{l|}{} &
 \multicolumn{4}{c}{Signal efficiency  error (\%) }    \\
 \multicolumn{2}{l|}{} &
 \multicolumn{4}{c}{Event Selection $\WW\rightarrow$ } \\
     &Source of uncertainty            &\qqen\ & \qqmn\ & \qqtn\ & \qqln \\ \hline\hline
i)   &MC Statistics                    & 0.07  & 0.06 & 0.10 & 0.04 \\     
ii)  &WW Fragmentation                 & 0.25  & 0.20 & 0.50 & 0.20 \\ 
iii) &Tau candidate ID                 & $-$   & $-$  & 0.60 & 0.20 \\
iv)  &$\cal{O}(\alpha)$ QED/Electroweak& 0.09  & 0.05 & 0.03 & 0.04 \\
 v)  &ISR and FSR                      & 0.07  & 0.12 & 0.10 & 0.03 \\ \hline
vi)  &ECAL energy response             & 0.11  & $-$  & 0.08 & 0.03 \\
vii) &Track momentum response          & 0.07  & 0.05 & 0.08 & 0.02 \\
viii)&Jet energy response              & 0.01  & $-$  & 0.02 & 0.01 \\ 
ix)  &Tracking Losses                  & 0.30  & 0.05 & 0.06 & 0.10 \\
x)   &Detector Occupancy               & 0.03  & 0.03 & 0.06 & 0.03 \\ \hline
xi)  &Preselection                     & 0.10  & 0.10 & 0.15 & 0.12 \\
xii) &Likelihood Selection             & 0.30  & 0.10 & 0.40 & 0.10 \\
     & Other                           & 0.04  & 0.03 & 0.02 & 0.03 \\ \hline
     &Total                            & 0.54  & 0.30 & 0.91 & 0.36 \\ \hline\hline
 \end{tabular}
 \end{center}
 \caption{ Sources of uncertainty on the \WWqqln\
           selection efficiencies. The errors quoted apply to the selection
           efficiency for the combined $\roots=183-209$\,GeV data set. 
           Entries where the systematic error estimate is less than
           0.01\,\% are denoted by $-$. The errors on the combined
           $\qqlv$ selection take into account correlations 
           between the separate channels.
 \label{tab:qqln_effsys}}
\end{table}

Table~\ref{tab:qqln_effsys} lists the various contributions to the 
systematic uncertainty on the $\qqen$, $\qqmn$ and $\qqtn$ selection 
efficiencies. Many of the potential systematic effects primarily 
affect the classification of selected $\qqln$ events rather than
the overall $\qqln$ efficiency. Amongst 
the effects studied were: 

\medskip\noindent
{\bf i)} Finite MC statistics of the 
\KANDY\ MC samples used to determine the efficiencies.

\medskip\noindent
{\bf ii)}
The fragmentation and hadronisation systematic uncertainties are studied with
fully simulated MC \WWqqln\ samples where the hadronisation
process is modelled using \JETSET, \HERWIG\ or \ARIADNE.
In addition, the parameters $\sigma_q$, $b$, $\Lambda_{QCD}$, and $Q_{0}$ 
of the \JETSET\ fragmentation model are varied by one standard 
deviation about their tuned values~\cite{bib:qqqqQCD}.

\medskip\noindent
{\bf iii)} 
The largest single systematic uncertainty in the $\qqln$ selection
is due to an identified deficiency in the MC simulation of
isolated tracks from the fragmentation/hadronisation process. 
Such tracks, if sufficiently isolated can have similar properties 
to those from hadronic tau decays. In data there is a clear excess of 
low momentum tracks which have been identified as the best tau decay
candidate compared to the MC expectation. This excess persists at all 
stages in the event selection; for example, there is a $\sim$10\,\% excess of 
data events passing the $\WWqqtn$ preselection cuts 
(a sample dominated by background from $\eetoqq$). 
To assess the impact on the $\qqtn$ analysis, a control sample  
of two jet events is formed by removing the tracks and calorimeter clusters 
associated with the lepton in selected 
$\qqen$ and $\qqmn$ events. The full $\qqtn$ event selection is
applied to these events and the selection efficiency is found to be 
$7.3\pm4.6\,\%$ higher in data than the MC expectation. Again there is a clear 
excess ($25\pm 7\,\%$) of isolated tracks with momenta less than 5\,GeV. 
This data sample is used to provide a momentum dependent correction 
factor which is used to reweight all MC events where a fragmentation track 
is identified as the best tau candidate. After applying this correction, 
the data/MC agreement at all stages in the $\qqtn$ selection is significantly
improved. The effect of this correction is to increase the expected background from 
$\qqnn$ and single-W ($\Wenu$) events. Because $\qqtn$ events can
also be selected on the basis of a fragmentation track, the predicted 
selection efficiency for $\qqtn$ events is also increased by 0.6\,\%.
The full size of the corrections to efficiency and background 
are assigned as (correlated) systematic errors in the $\qqtn$ selection.

\medskip\noindent
{\bf iv)} The
selection efficiencies are sensitive to hard photon radiation
in the W-pair production process. The \OPAL\ data are consistent
with the predictions from \KANDY~\cite{bib:opalwwg}.
Potential systematic biases are
estimated by reweighting the \KANDY\ MC samples so as to turn off
the $\cal{O}(\alpha)$ electroweak treatment of radiation from the 
W-bosons.

\medskip\noindent
{\bf v)} A conservative estimate of the possible biases
arising from FSR from the lepton or tau decay products 
is investigated by reweighting the MC so as to
change the rate of such FSR by $\pm50\,\%$. This mainly affects the 
classification of selected events.
The selection efficiencies are found to be insensitive 
to the detailed treatment of ISR. 

\medskip\noindent
{\bf vi), vii) and viii)} Uncertainties
in the detector calibration, linearity of energy response and 
MC simulation of the energy resolution were studied in
detail for the \OPAL\ analysis of the W-boson mass\cite{bib:opalmw}.
The uncertainties related to ECAL energy, track momentum and
jet energy response described therein are propagated to the 
event selection.

\medskip\noindent
{\bf ix)} $\Zz\rightarrow\ell^+\ell^-$ events
are used to study the tracking efficiency for electrons and muons.
It is found that the MC overestimates the efficiency for reconstructing
electron and muon tracks in the forward region, 
$|\cos\theta|>0.9$. The effect on the selection efficiency is reduced 
by a factor of approximately three due to the trackless selections.
The MC efficiency estimates are corrected and the full size of the correction 
is assigned as a systematic error.

\medskip\noindent
{\bf x)} Randomly triggered
events recorded throughout the data-taking period 
are used to assess the impact of energy deposits in the 
detector (particularly in the forward luminosity calorimeters) 
which can result in the event being vetoed. As a result, the
MC efficiencies were corrected and half the correction
assigned as a systematic uncertainty. 

\medskip\noindent
{\bf xi)} The event 
preselection cuts remove approximately $1\,\%$ of $\qqln$ events.
Possible systematic effects specifically associated with the preselection
(in addition to those described above) are studied applying the
likelihood selection to all events failing just one of the preselection
cuts. There is no evidence of any systematic bias and the statistical
precision of the study is used to assign the systematic uncertainty.

\medskip\noindent
{\bf xii)} The MC expectation for each of the 
variables used in the likelihood selection is compared to the observed
distribution for the selected events. The ratio of data to MC 
is used to define bin-by-bin corrections for each distribution. These 
corrections are propagated back into the likelihood selection and the 
associated
systematic errors are obtained from the resulting changes in the 
selection efficiencies.

\medskip\noindent
{\bf Background Uncertainties:}
Table~\ref{tab:backgrounds} shows the background cross sections and
total uncertainties for the three \qqlns\ selections. 
The largest contributions to
the background in the $\qqlns$ selections are from the four
fermion final states $\qqen$, $\qqll$ and $\qqnn$ and from $\eetoqq$.
In the $\qqtn$ selection, the uncertainties on the four fermion 
backgrounds are dominated by the correction for isolated low 
momentum tracks described above. The $\qqen$ background mainly 
arises from the single W process (including interference with the \CC\ 
diagrams); a 5\,\% uncertainty on this cross section is assumed\cite{bib:lep4f}. 
Background from the $\eetoqq$ process mainly arises from radiative
return events with an unobserved photon in the beam direction where
a hadronisation track is mis-identified as the lepton.
The $\eetoqq$ background is assigned a 10\,\% systematic uncertainty
for the MC modelling of the hadronisation process (based on
comparisons of \PYTHIA, \HERWIG\ and \ARIADNE). 
The MC estimate of this background rate is checked using 
control samples constructed from the data directly.
For the background, `fake' events are constructed by boosting
hadronic \Zz\ events recorded at $\roots = 91$~GeV to the
invariant mass distribution expected of quark pairs
at the appropriate $\roots$.
There is an additional $11\,\%$ uncertainty on the $\eetoqq$ 
background in the $\qqen$ selection from uncertainties in the rate
at which high energy photon conversions fake an electron. 
The backgrounds from multi-peripheral two photon processes
(almost entirely from hadronic final states rather
than from $\epem\rightarrow\epem\lplm$) are assigned a systematic 
uncertainty of $50\,\%$ to cover the variation in predictions obtained
from different generators. 

\subsubsection{\boldmath $\WWqqln$ Results}

Using the \KANDY\ MC samples the inclusive \qqlns\ selection 
is estimated to be $83.8 \pm 0.4\,\%$ efficient for \WWqqln\ 
events. The selection efficiencies for the different centre-of-mass 
energies are listed in Table~\ref{tab:qqlv_xs}. Above the 
\WW\ threshold region the selection efficiency does not depend
strongly on the centre-of-mass energy. The luminosity weighted efficiencies of 
the \WWqqln\ 
selection for the individual channels are given in 
Table~\ref{tab:lumi_weighted_effic}. 
The efficiencies/numbers of expected events
in all tables include small corrections ($0.1-0.3\,\%$) 
which account for tracking losses which are not modelled by
the MC simulation of the OPAL detector. The effect of detector 
occupancy from beam-related backgrounds is also included as is the
small correction associated with the identification of tau candidates
described above. 
  
\begin{figure}
 \begin{center}
   \epsfig{file=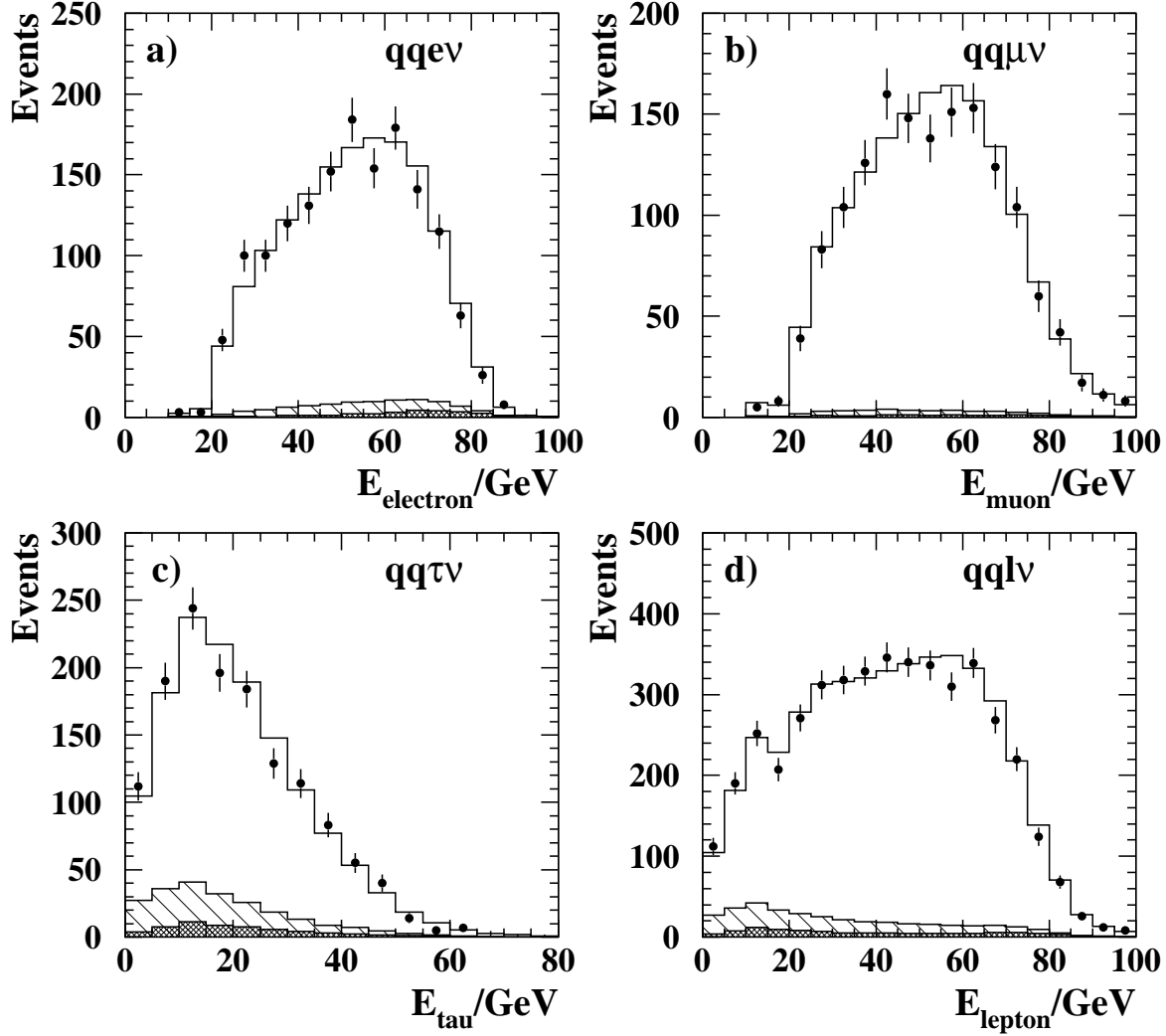,width=\textwidth}
   \caption{ 
     Distributions of measured energies of the electrons, muons and visible tau decay products
     for events selected as
     \qqev, \qqmv, and \qqtv\ respectively. 
     The combined distribution for all events selected 
     as \qqlv\ is also shown. 
     The data are shown as the points with statistical error bars, while
     the histogram is the total MC expectation.
     The combined background from two-fermion and two-photon processes 
     is shown by the cross-hatched
     region, while the non-\CC\ four-fermion background is shown by the
     single-hatched region.
     \label{fig:qqlnsel}
     }
 \end{center}
\end{figure}

In total 4572 events are selected as inclusive \WWqqln\ candidates in
agreement with the SM expectation of $4622\pm28$. 
Figure~\ref{fig:qqlnsel} shows distributions of the reconstructed 
energy of the lepton in the \qqen, \qqmn, and \qqtn\ selections
and the summed distribution. The data distributions are in good 
agreement with the MC expectations. 

The numbers of selected $\qqln$ events at each energy are used to
determine the cross sections for $\epem\rightarrow\WW\rightarrow\qqln$
given in Table~\ref{tab:qqlv_xs}. The results are obtained assuming
the small backgrounds from $\lnln$ and $\qqqq$ are given by the SM. 
The measured cross sections are in agreement with the SM expectations.

 \begin{table}[htbp]
   \centering
    \begin{tabular}{c|cccc|c|c} \hline\hline
  $\roots$ & $\cal{L}$ & $N$ & Efficiency & Background     & $\sigma(\WW\rightarrow\qqlv)$ & SM \\
   $[\mathrm{GeV}]$    & [pb$^{-1}$]  & [events]  & [\%]   & [events] & [pb] & [pb]\\ \hline\hline
161.30&  9.9&   12& 63.6$\pm$  2.5&  1.4$\pm$  0.5& 1.68$\pm$ 0.55$\pm$ 0.07& 1.58\\
172.11& 10.4&   55& 84.2$\pm$  1.0&  4.6$\pm$  0.8& 5.77$\pm$ 0.85$\pm$ 0.07& 5.31\\ \hline
182.68& 57.4&  357& 84.2$\pm$  0.4& 22.1$\pm$  2.1& 6.93$\pm$ 0.39$\pm$ 0.05& 6.74\\
188.63&183.0& 1171& 84.6$\pm$  0.4& 89.8$\pm$  5.7& 6.98$\pm$ 0.22$\pm$ 0.05& 7.13\\
191.61& 29.3&  176& 84.6$\pm$  0.4& 15.1$\pm$  1.0& 6.48$\pm$ 0.54$\pm$ 0.05& 7.26\\
195.54& 76.4&  554& 84.1$\pm$  0.4& 43.6$\pm$  2.6& 7.94$\pm$ 0.37$\pm$ 0.05& 7.38\\
199.54& 76.6&  494& 83.7$\pm$  0.4& 44.8$\pm$  2.7& 7.01$\pm$ 0.35$\pm$ 0.05& 7.46\\
201.65& 37.7&  255& 83.6$\pm$  0.4& 22.1$\pm$  1.3& 7.39$\pm$ 0.51$\pm$ 0.05& 7.48\\
204.88& 81.9&  523& 83.9$\pm$  0.4& 52.3$\pm$  3.2& 6.85$\pm$ 0.33$\pm$ 0.05& 7.50\\
206.56&138.5&  975& 83.6$\pm$  0.4& 86.9$\pm$  5.1& 7.67$\pm$ 0.27$\pm$ 0.05& 7.51\\
 \hline\hline
 \end{tabular}
 \caption{\label{tab:qqlv_xs} Measured cross sections for
 the process $\epem\rightarrow\WW\rightarrow\qqlv$.
 For the $\qqlv$ selection the data below $\roots=182.68$\,GeV have not been reanalysed and the 
 results are taken from~\cite{bib:xs172,bib:xs161}. 
 The errors on the cross sections are statistical
 and systematic respectively.
 The numbers of selected events,
 $\qqlv$ selection efficiencies and expected numbers of
 background events are also listed.
 The backgrounds include fully-leptonic and
 fully-hadronic $\WW$ decays for which the cross sections
 are taken to be their SM expectations.
 }
 \end{table}

\subsection{\boldmath Selection of $\WWqqqq$ events}

\label{sec:selqqqq} 

The selection of fully hadronic \WWqqqq\ events is performed in two 
stages using a cut-based preselection followed by a likelihood
selection procedure. 
This likelihood selection is primarily designed to reject the dominant 
background from the $\eetoqq$ process where the di-quark system 
fragments into a four jet topology.
No attempt is made to discriminate against the neutral current 
process \ZZqqqq\ for which the cross section is at least an order of
magnitude smaller than that for \WWqqqq.
The preselection and likelihood selection variables are unchanged
from those described in previous OPAL publications~\cite{bib:xs189}
although the tuning of the likelihood discriminant is updated for
different ranges of $\sqrt{s}$.

\subsubsection{Event Selection}

All events which are classified as hadronic~\cite{bib:tkmh} and which 
have not been selected by either the \lnlns\ or the \qqlns\ selections 
are considered as candidates for the \WWqqqq\ selection.
In addition, any event which is identified and rejected as a four-fermion background
event in the \qqlns\ selection is also rejected as a \qqqq\ candidate 
event.

Tracks and calorimeter clusters are combined into four jets using the
Durham algorithm~\cite{bib:Durham} and the total momentum and energy of 
each jet is corrected for double-counting of energy~\cite{bib:GCE}.
To remove events which are clearly inconsistent with a fully hadronic 
\WW\ decay, candidate events are required to satisfy a set of 
preselection cuts including a cut on minimum visible energy (70\,\% of
\roots), minimum invariant mass (75\,\% of \roots), and minimum 
multiplicity per jet (one track).
The most important preselection cut is 
$\log_{10}(\WQCD) < 0$~\cite{bib:qcd420}, where
\WQCD\ is the QCD matrix element calculated 
as an event weight formed from the tree level ${\cal O}(\alpha_s^2)$ 
matrix element~\cite{bib:ERT} for the four jet production processes 
($\eetoqq\rightarrow\qqqq,\qqgg$). 
The value of \WQCD\ is determined by using the observed 
momenta of the four reconstructed jets as estimates of the underlying
parton momenta which are input to the matrix element calculation.
The best discriminating power between signal and background was found 
using a variable defined as the largest value of the $\WQCD$ 
matrix element from any of the 24 possible jet-parton associations 
in each event.

The preselection requirements reject around $95$\,\% of the \eetoqq\ 
events which comprise the dominant source of background in the 
\WWqqqq\ event selection, while
the preselection efficiency for the hadronic \WWqqqq\ decays 
is estimated to be $90-93$\,\% depending on $\roots$.

\begin{figure}
 \begin{center}
   \epsfig{file=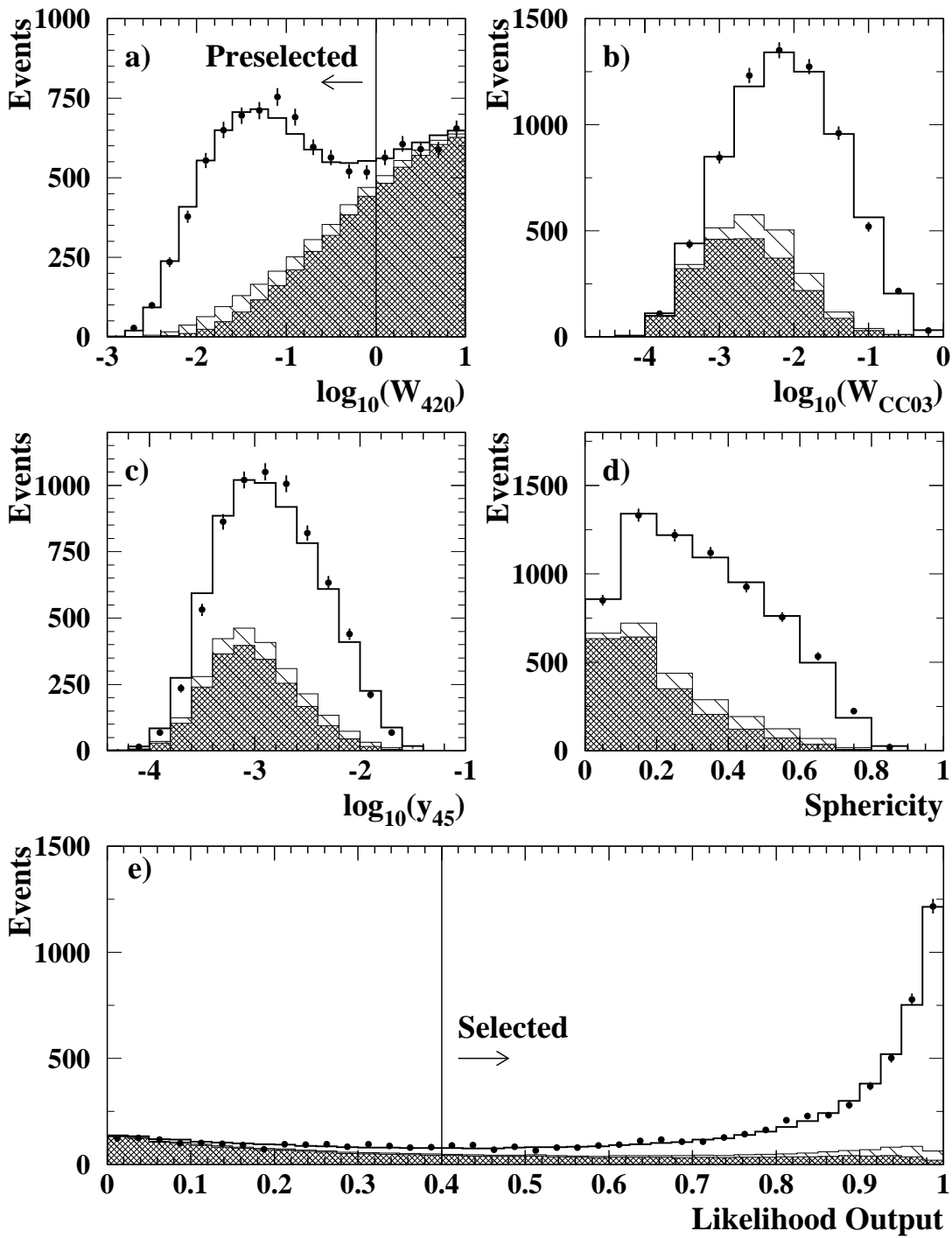,width=0.94\textwidth}
   \caption{ 
     \label{fig:qqqqsel}
Distributions of the variables (described in the text) used in the likelihood 
selection of 
$\WWqqqq$ events (a)-(d) and the resulting relative likelihood distribution (e).
All plots are shown for the combined sample from
data recorded between $\sqrt{s}=183-209$\,GeV.
The data are shown as the points with error bars (statistical errors only).
The total Standard Model MC prediction is shown by the unshaded histogram.
The background components are also shown: four-fermion background 
(singly-hatched) and two-fermion background (cross-hatched).
The MC is normalised to the integrated luminosity of the data. 
     }
 \end{center}
\end{figure}

Events satisfying the preselection cuts are classified as signal or
background based upon a four variable likelihood selection.
The following likelihood variables are selected to provide a good 
separation between the hadronic \WWqqqq\ signal and the \eetoqq\ four 
jet background, while minimising the total number of variables used:
\begin{itemize}
        \item  $\log_{10}(\WQCD)$, the QCD four jet matrix element;
        \item  $\log_{10}(\WCC)$, the \EXCALIBUR\ matrix 
               element~\cite{bib:EXCALIBUR} for the \CC\ process (\WWqqqq);
        \item  $\log_{10}(y_{45})$, the logarithm of the value of the Durham 
               jet resolution parameter at which an event is reclassified from 
               four jets to five jets;
        \item  event sphericity.
\end{itemize}
Figure~\ref{fig:qqqqsel} shows the distribution of these four likelihood 
variables for all preselected events found in the $183-209$\,GeV data.
To improve the statistical power of this selection, a multi-dimensional
likelihood technique is used to account for the correlations between 
the four likelihood input variables~\cite{bib:PC}.
Most of the separation between the signal and background events is
provided by the two matrix element values $\log_{10}(\WCC)$ and
$\log_{10}(\WQCD)$, which is related to the relative probability that the 
kinematics of the observed event are consistent with signal or 
background production respectively.
While the likelihood input variables are the same for events in
all $\roots$ ranges, the likelihood discriminant functions are
separately calculated from \CC\ signal and \eetoqq\ background 
MC samples in three ranges of $\roots$:  $185-194$\,GeV, $194-202.5$\,GeV,
and $202.5-209.0\,$GeV.
Candidate events at $\roots$ below 185~GeV are unchanged from previous
OPAL publications~\cite{bib:xs172,bib:xs183,bib:xs161}.

An event is selected as a hadronic \WWqqqq\ candidate if the likelihood
discriminant variable, also shown in Figure
\ref{fig:qqqqsel}, is greater than 0.4.
This cut value was chosen to maximise the expected statistical power 
of this selection assuming the SM rate for \CC\ production.

\subsubsection{Background Estimation}

The accepted \eetoqq\ background is estimated from \KK\ MC 
samples, with \PYTHIA\, \HERWIG\ and \ARIADNE\ hadronisation 
being used as cross-checks.
To reduce the uncertainty on this background estimate, a technique 
to measure this rate directly from the data is used.
By comparing the number of events seen in data and MC 
in the range $0<\log_{10}(\WQCD)<1$ which would otherwise pass the
preselection cuts, the overall four jet background rate predicted 
by the MC is normalised to the observed data.
This procedure is performed and applied separately in the three 
$\roots$ selection ranges described above.
A luminosity-weighted average correction over the full
$\roots$ range of $(-1.4 \pm 1.7)\,\%$ is found for the default
\KK\ samples, where the uncertainty is the statistical precision
of the normalisation procedure.
The observed data and corrected MC expectation in this 
sideband background region are shown in Figure~\ref{fig:qqqqsel}.
The expected contamination from \CC\ production in this region is 
less than 3\,\%, resulting in a negligible bias on the extracted \CC\ 
cross section.

\subsubsection{Selection Uncertainties}

The main systematic uncertainty on the selection efficiency
results from the modelling of the QCD hadronisation process.
This uncertainty is estimated by comparing the selection
efficiency predicted using the \JETSET\ hadronisation model 
with alternative models including \HERWIG, \ARIADNE\ and an
older version of the OPAL \JETSET\ tuning \cite{bib:oldjetset}.
These variations cover the observed data/MC differences such
as the $y_{45}$ distribution shown in Figure~\ref{fig:qqqqsel}. 
The uncertainty in the selection efficiency from the modelling
of the hadronisation process is almost exclusively due to the
preselection requirements, and is found to be independent of
$\roots$.
The largest observed deviation in selection efficiency is taken
as the systematic uncertainty, resulting in an estimated relative uncertainty
of 0.9\,\% which is fully correlated
between different $\roots$ samples.

Cross-checks of this uncertainty are performed by comparing
the observed shapes of both the preselection and selection variables
seen in data to those predicted by the signal MC samples.
After subtracting the expected background, the differences between
observed data and expected MC signal distributions are comparable
to the variations observed within the different hadronisation 
models themselves.  
In addition, the effect of directly varying the parameters
$\sigma_{\mathrm{q}}$, $b$, $\Lambda_{\mathrm{QCD}}$, and $Q_{0}$ 
of the \JETSET\ hadronisation model by one standard deviation 
about their tuned values~\cite{bib:qqqqQCD} as was done for
previous OPAL results~\cite{bib:xs189} leads to similar 
uncertainties.

Additional uncertainties on the modelling of the underlying
hard process are evaluated by comparing 
\CC\ events produced by \KANDY\ with
other generators (\EXCALIBUR, \PYTHIA, and \GRC~\cite{bib:grc4f}).
Uncertainties on the detector modelling are evaluated from 
direct comparison of data distributions with MC predictions,
and are generally smaller than the observed differences seen
between the different hadronisation models.
Possible biases related to final state interactions between the
hadronic systems produced by different W bosons have been evaluated
for colour-reconnection effects~\cite{bib:CR} and Bose-Einstein 
correlations~\cite{bib:BE}.
These effects are found to be small, and the total change in 
predicted selection efficiency when these effects are included
in the hadronisation model is taken as the systematic uncertainty.

\subsubsection{Background Uncertainties}

The dominant uncertainty on the expected background rate comes from
the modelling of the hadronisation process, particularly in \eetoqq\
events.
This uncertainty is evaluated in the same manner as the hadronisation
uncertainty for the signal efficiency, using large MC samples 
produced with a variety of hadronisation models, and taking the
largest observed deviation as an estimate of the systematic uncertainty.
The background normalisation procedure has been consistently 
applied during these systematic checks.
The uncertainty on the estimated background is about 75~fb (the exact
value depends on the centre-of-mass energy) which is taken to be 
fully correlated between different $\roots$ samples.
The uncertainty from modelling of the hadronisation process for 
the background estimation is found to be largely uncorrelated with 
the uncertainty on the signal efficiency. 

The background normalisation procedure contributes an additional,
statistical uncertainty to the background estimation of about 3\,\%
which is uncorrelated between different $\roots$ ranges.
Additional uncertainties in the non-\CC\ four-fermion background are 
estimated by comparing the expectations of \KORALW, \GRC, 
and \EXCALIBUR.
This background is predominantly from the neutral current process 
\ZZqqqq, of which only 20\,\% is in final states with direct 
interference with the \CC\ diagrams.
In each case, the single largest difference observed in a set of 
systematic checks is taken as an estimate of the uncertainty.

\subsubsection{\boldmath $\WWqqqq$ Results}
The luminosity-weighted efficiency of the likelihood selection for 
\WWqqqq\ events is estimated from \KANDY\ MC samples to be 
$85.9 \pm 0.9\,\%$, 
where the error represents an estimate of the systematic uncertainties.
A total of 5933 \WWqqqq\ candidate events are selected compared to
the expectation of $5845.2\pm67.5$. The  luminosity-weighted purity of the 
selected event sample is 77\,\%. 
The selection efficiencies for the different centre-of-mass energies
are listed in Table~\ref{tab:qqqq_xs}. 
For the 189$-$209\,GeV data the selection efficiency does not depend strongly
of centre-of-mass energy. 
The numbers of selected $\qqqq$ events at 
each energy are used to determine cross sections for 
$\epem\rightarrow\WW\rightarrow\qqqq$, also listed in 
Table~\ref{tab:qqqq_xs}. The results are obtained assuming
the small backgrounds from $\lnln$ and $\qqln$ are given by the SM. 
The measured cross sections are in agreement with the SM expectations.

 \begin{table}[htbp]
   \centering
    \begin{tabular}{c|cccc|c|c} \hline\hline
  $\roots$ & $\cal{L}$ & $N$ & Efficiency & Background     & $\sigma(\WW\rightarrow\qqqq)$& SM \\
   $[\mathrm{GeV}]$    & [pb$^{-1}$]  & [events]  & [\%]   & [events] & [pb] & [pb]\\ \hline\hline
161.30&  9.9&   14& 56.7$\pm$  3.5&  3.4$\pm$  0.4& 1.88$\pm$ 0.67$\pm$ 0.14& 1.64\\
172.11& 10.4&   54& 70.3$\pm$  3.0& 13.1$\pm$  1.9& 5.62$\pm$ 1.01$\pm$ 0.24& 5.52\\ \hline
182.68& 57.4&  439& 86.3$\pm$  0.9& 98.1$\pm$  6.8& 6.89$\pm$ 0.42$\pm$ 0.11& 7.00\\
188.63&183.0& 1553& 86.6$\pm$  0.9&339.5$\pm$ 17.8& 7.66$\pm$ 0.25$\pm$ 0.12& 7.41\\
191.61& 29.3&  245& 86.2$\pm$  0.9& 55.2$\pm$  2.8& 7.51$\pm$ 0.62$\pm$ 0.12& 7.54\\
195.54& 76.4&  709& 87.2$\pm$  0.9&152.6$\pm$  7.8& 8.35$\pm$ 0.40$\pm$ 0.12& 7.67\\
199.54& 76.6&  643& 86.7$\pm$  0.9&150.6$\pm$  7.7& 7.42$\pm$ 0.38$\pm$ 0.11& 7.75\\
201.65& 37.7&  342& 86.6$\pm$  0.9& 75.8$\pm$  3.8& 8.16$\pm$ 0.57$\pm$ 0.12& 7.77\\
204.88& 81.9&  683& 86.3$\pm$  0.9&159.9$\pm$  8.2& 7.40$\pm$ 0.37$\pm$ 0.11& 7.79\\
206.56&138.5& 1251& 86.1$\pm$  0.9&274.4$\pm$ 13.9& 8.19$\pm$ 0.30$\pm$ 0.12& 7.80\\
 \hline\hline
 \end{tabular}
 \caption{\label{tab:qqqq_xs} Measured cross sections for
 the process $\epem\rightarrow\WW\rightarrow\qqqq$.
 For the $\qqqq$ selection the data below $\roots=182.68$\,GeV have not been reanalysed and the 
 results are taken from~\cite{bib:xs172,bib:xs161}. 
 The errors on the cross sections are statistical
 and systematic respectively.
 The numbers of selected events,
 $\qqqq$ selection efficiencies and expected numbers of
 background events are also listed.
 The backgrounds include fully-leptonic and
 semi-leptonic $\WW$ decays which for the cross sections
 are taken to be fixed to their SM expectations.
 }
 \end{table}

\section{\boldmath Measurement of the \WW\ cross section}

The observed numbers of selected \WW\ events are used to measure the
\WW\ production cross section and the W decay branching fractions 
to leptons and hadrons.  
The measured cross section corresponds to that of W-pair production 
from the \CC\ diagrams as discussed earlier.
The expected four-fermion backgrounds quoted throughout this 
paper include contributions from both non-\CC\ final states and 
the effects of interference with the \CC\ diagrams.  
Mis-identified \CC\ final states are not included in the background
values listed in Table~\ref{tab:backgrounds}, but rather are
taken into account by off-diagonal entries in the efficiency matrix.
Table~\ref{tab:selection_summaries} summarises the event selections in the
ten \WW\ decay topologies.

 \begin{table}[htbp]
   \centering
    \begin{tabular}{c|cc|cc|c} \hline\hline
  Selection & Efficiency & Purity & Expected & Observed & Data/Expected \\ \hline\hline
\Sevev& 89.0\,\% & 88.1\,\% & 136.7$\pm$  2.4&   141&1.032$\pm$0.087$\pm$0.018\\
\Smvmv& 95.0\,\% & 89.9\,\% & 143.0$\pm$  2.5&   156&1.091$\pm$0.087$\pm$0.017\\
\Stvtv& 71.8\,\% & 79.5\,\% & 122.2$\pm$  3.4&   131&1.072$\pm$0.094$\pm$0.028\\
\Sevmv& 91.8\,\% & 93.9\,\% & 264.8$\pm$  3.2&   251&0.948$\pm$0.060$\pm$0.012\\
\Sevtv& 81.9\,\% & 88.5\,\% & 250.5$\pm$  4.2&   256&1.022$\pm$0.064$\pm$0.017\\
\Smvtv& 75.6\,\% & 92.6\,\% & 220.9$\pm$  4.1&   253&1.145$\pm$0.072$\pm$0.019\\
 \hline
\Slvlv& 83.8\,\% & 89.7\,\% &1137.7$\pm$  8.5&  1188&1.044$\pm$0.030$\pm$0.007\\
 \hline
\Sqqev& 88.3\,\% & 93.2\,\% &1597.5$\pm$  9.8&  1585&0.992$\pm$0.025$\pm$0.006\\
\Sqqmv& 92.8\,\% & 96.8\,\% &1616.7$\pm$  5.1&  1581&0.978$\pm$0.025$\pm$0.003\\
\Sqqtv& 70.1\,\% & 84.1\,\% &1407.8$\pm$ 23.6&  1406&0.999$\pm$0.027$\pm$0.017\\
 \hline
\Sqqlv& 83.8\,\% & 91.7\,\% &4622.0$\pm$ 27.6&  4572&0.989$\pm$0.015$\pm$0.006\\
 \hline
\Sqqqq& 85.9\,\% & 77.4\,\% & 5845.2$\pm$ 67.5&  5933&1.015$\pm$0.013$\pm$0.012\\ \hline
Total& 85.2\,\% & 84.7\,\% &11604.8$\pm$ 73.4& 11693&1.008$\pm$0.009$\pm$0.006\\
 \hline\hline
 \end{tabular}
 \caption{\label{tab:selection_summaries}
 Selected events in the each of the 10 $\WW$ decay
 topologies compared to the SM expectation.
 Also listed are the combined numbers for the six
 $\lvlv$ decay channels and for the three $\qqlv$ decay channels.
 The efficiencies and purities for the $\lvlv$ ($\qqlv$) decay
 channels are calculated treating all $\lvlv$ ($\qqlv$)
  events as signal; e.g. the quoted efficiencies in the
  $\lvlv$ channels represent the selected \CC\ cross 
  section for any $\lvlv$ flavour divided by the generated
  \CC\ cross section in the specific channel. Note that the 
  total ratio of data to MC is for the sum of signal 
  and background events.
 }
 \end{table}

\begin{figure}
 \begin{center}
   \epsfig{file=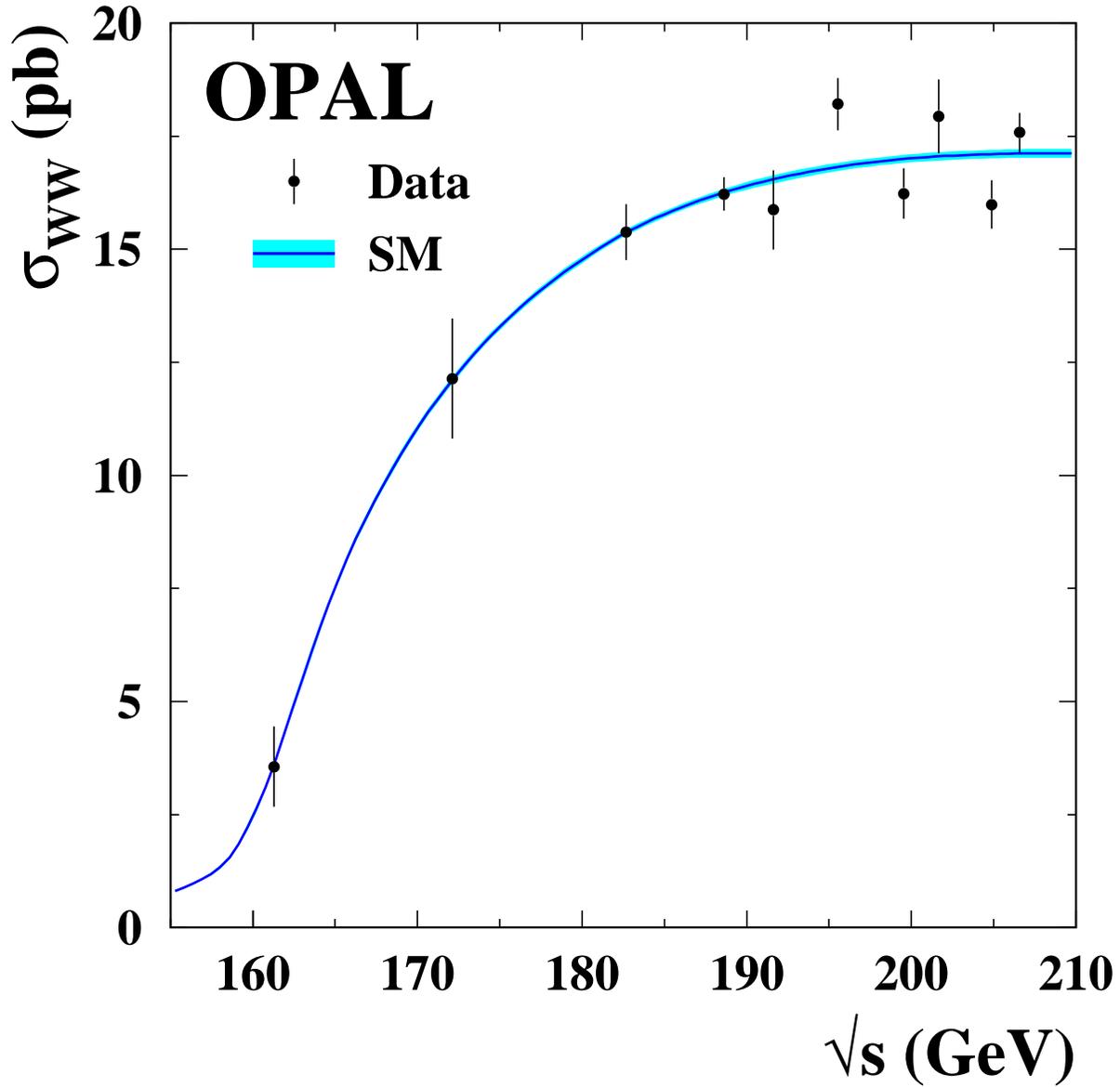,width=\textwidth}
   \caption{ 
     The measured WW cross sections from fits assuming SM W decay 
     branching fractions.
     The measured cross sections (points) are compared to the
     SM expectation (line) which is the average of the predictions
     from \YFSWW\ and \RACWW. The shaded region shows the 0.5\,\%
     theoretical error. 
     \label{fig:xsec}
     }
 \end{center}
\end{figure}

The \WW\ cross section and branching 
fractions are measured using data from the ten separate decay channels. 
The physical parameters (cross sections, branching ratios, etc.) 
are obtained from fits where all correlated systematic
uncertainties are taken into account. The total cross section is obtained 
from a maximum likelihood fit to the numbers of events in the ten decay channels 
from data at all centre-of-mass energies
allowing the cross sections at each centre-of-mass
energy to vary and assuming the SM branching fractions. 
Efficiency, background, and luminosity systematic uncertainties are included 
as nuisance parameters with Gaussian penalty terms in the likelihood 
function\cite{bib:pdglike}. 
Correlations are accounted for in the covariance matrix of the nuisance 
parameters associated with the systematic uncertainties. 
The results are listed in Table~\ref{tab:xsec_combined_fit} and shown
in Figure~\ref{fig:xsec}. In both cases the results are 
compared to the SM expectation which is taken to be the mean of the
cross sections predicted by \YFSWW\ and \RACWW\ (on average the predicted
cross section from \YFSWW\ is 0.2\,\% higher than that from \RACWW). 
The results do not differ significantly if the 
SM branching fractions are left unconstrained in the fit. When compared to
the SM expectations, the 10 cross section measurements in Figure~\ref{fig:xsec}
yield a $\chi^2$ of 15.5 (11\,\% probability). When the 100 individual event
counts used to obtain the cross sections (ten channels $\times$ ten $\roots$ bins)
are compared to the SM expectation the $\chi^2$ obtained is 94.5 for 100 degrees
of freedom. The \OPAL\ \WW\ data are consistent with
the SM expectation.
The cross sections listed in Table~\ref{tab:xsec_combined_fit} differ from
than the sums of the exclusive cross sections from the separate channels
(listed in Tables~\ref{tab:lvlv_xs}, \ref{tab:qqlv_xs} and \ref{tab:qqqq_xs}) because
of the constraint to the SM branching ratios and the larger systematic
errors and in the $\qqqq$ channel.

 \begin{table}[htbp]
   \centering
    \begin{tabular}{c|c|c} \hline\hline
  $\langle\roots\rangle$/GeV  & \sigWW\ [pb] & \sigWWSM\ [pb] \\ \hline
161.30&  3.56$\pm$  0.88$\pm$  0.11&  3.61\\
172.11& 12.14$\pm$  1.34$\pm$  0.22& 12.10\\
182.68& 15.38$\pm$  0.61$\pm$  0.13& 15.37\\
188.63& 16.22$\pm$  0.35$\pm$  0.11& 16.26\\
191.61& 15.87$\pm$  0.86$\pm$  0.10& 16.55\\
195.54& 18.21$\pm$  0.57$\pm$  0.12& 16.82\\
199.54& 16.23$\pm$  0.54$\pm$  0.11& 17.00\\
201.65& 17.94$\pm$  0.81$\pm$  0.11& 17.05\\
204.88& 15.99$\pm$  0.52$\pm$  0.11& 17.10\\
206.56& 17.58$\pm$  0.42$\pm$  0.12& 17.12\\
 \hline\hline
 \end{tabular}
 \caption{Measured \CC\ \WW\ cross sections from a combined fit to
  all data. The last column shows the SM expectations 
  which are taken from the average of the predictions from
  \YFSWW\ and \RACWW.
 \label{tab:xsec_combined_fit} }
 \end{table}

A fit to the data where the expected cross sections at all 
centre-of-mass energies are given by the
SM expectation scaled by a single data/SM ratio gives:
$$ \mathrm{data}/\mathrm{SM} = \rsmresult,$$
where the SM expectation is the mean of the
cross sections predicted by \YFSWW\ and \RACWW.

\label{sec:wwxsec}

\section{\boldmath Measurement of the W Branching Fractions}

\label{sec:wwbr}

A simultaneous fit to the numbers of \WW\ candidate events in the 
ten identified final states (\enens, \mnmns, \tntns, \enmns, \entns, 
\mntns, \qqen, \qqmn, \qqtn, and \qqqq) observed by OPAL at each of
the ten centre-of-mass energies between 161\,GeV and 207\,GeV 
gives the following values for the leptonic branching fractions 
of the W boson:
\begin{eqnarray*}
  \Br(\Wtoen) & = &  \brevresult  \\ 
  \Br(\Wtomn) & = &  \brmvresult  \\ 
  \Br(\Wtotn) & = &  \brtvresult.
\end{eqnarray*}
Correlations between the systematic uncertainties at the different
energy points have been accounted for in the fit as have correlations
in the selection efficiency uncertainties for the different channels.
These results are consistent with the hypothesis of lepton universality, 
and agree well with the SM prediction of \SMbrlv\cite{bib:yellow}.
The correlation coefficient for the resulting values of 
$\Br(\Wtoen)$ and $\Br(\Wtomn)$ is $+0.14$. 
The correlation coefficients for 
$\Br(\Wtoen)$ and $\Br(\Wtomn)$ with the measurement of 
$\Br(\Wtotn)$ are $-0.30$ and $-0.23$ respectively.
A simultaneous fit assuming lepton universality gives
\begin{eqnarray*} 
  \Br(\Wtoqq) & = & \brqqresult,
\end{eqnarray*}
which is consistent with the SM expectation of \SMbrqq.
Here, the largest single source of systematic uncertainty is 
that from the \eetoqq\ background in the \WWqqqq\ channel.

Assuming the quark-lepton universality of the strength of the charged current
weak interaction, the hadronic branching fraction can be interpreted as a 
measurement of
the sum of the squares of the six elements of the CKM
mixing matrix, \Vij, which do not involve the top quark:
\begin{eqnarray*}
     \frac{\Br(\Wtoqq)}{(1-\Br(\Wtoqq))} & = &
 \left( 1+\frac{\alpha_s(\Mw)}{\pi} \right)
\sum_{i={\mathrm{u,c}}; \, j={\mathrm{d,s,b}}} \Vij^2.
\end{eqnarray*}
The theoretical uncertainty of this improved Born approximation 
due to missing higher order corrections is estimated to 
be 0.1\%~\cite{bib:yellow}.
Taking $\alpha_s(\Mw)$ to be $0.119\pm0.002$\cite{bib:pdg},
the branching fraction $\Br(\Wtoqq)$ from the $161-209$\,GeV
data yields
\begin{eqnarray*}
 \sum_{i={\mathrm{u,c}};\, j={\mathrm{d,s,b}}}
  \Vij^2 & = & \sumvresult,
\end{eqnarray*}
which is consistent with the value of 2 expected from unitarity in
a three-generation CKM matrix. 
If one assumes unitarity and a three-generation CKM matrix then this measurement 
can be interpreted 
as a test of quark-lepton universality of the weak coupling constant for
quarks, $\gW^{\qq}$, and for leptons, $\gW^{\lv}$:
\begin{eqnarray*}
 \gW^{\qq} / \gW^{\lv}= \gwratio.
\end{eqnarray*}

Finally, using the experimental measurements of the CKM 
matrix elements other than
$\Vcs$ gives
$\Vud^2+\Vus^2+\Vub^2+\Vcd^2+\Vcb^2 = 1.054\pm0.005$~\cite{bib:pdg},
and the \OPAL\ result for $\sum_{i={\mathrm{u,c}};\, j={\mathrm{d,s,b}}}
\Vij^2$ can be interpreted as a measurement of \Vcs\ which is the 
least well determined of these matrix elements:
\begin{eqnarray*}
  \Vcs & = & \vcsresult.
\end{eqnarray*}
The uncertainty in the sum of the other five CKM matrix elements,
which is dominated by the uncertainty on $\Vcd$, contributes
a negligible uncertainty of 0.003 to this determination of $\Vcs$.

\section{\bf \boldmath $\epem\rightarrow\WW$ Differential Cross Section}

\label{sec:wwdiffxsec}

In $\qqlv$ events it is possible to reconstruct the polar angle of 
the produced W$^-$ with respect to the $\mathrm{e}^-$ beam direction,
$\costw$, where the charge of the lepton tags the $\mathrm{W}^\pm$ 
and the jet momenta and the remaining event properties give the direction. 
Selected $\qqen$ and $\qqmn$ events are used to measure the 
differential cross section, $\mathrm{d}(\sigWW)/\mathrm{d}(\costw)$.
Events selected solely by the trackless selections are not used here.
Selected $\qqtn$ events are not considered due to the larger 
background and less reliable determination of lepton charge resulting from
the possibility of the candidate tau being formed from tracks 
from the fragmentation of the quarks.

The measured $\qqen$ and $\qqmn$ differential cross sections are corrected 
to correspond to the \CC\ set of diagrams but with the additional
constraint that, at generator level, the charged lepton is more than 
$20^\circ$ away from the 
$\epem$ beam direction, $20^\circ<\theta_{\ell^\pm}<160^\circ$. This angular
requirement is closely matched to the experimental acceptance. It also greatly 
reduces the difference between the full four-fermion cross section 
and the \CC\ 
cross section by reducing the contribution of $t$-channel single-W diagram in the 
$\qqen$ final state. At the MC generator level the angle $\costw$ is defined in terms of  
the four-momenta of the fermions from the ${\mathrm{W}^-}$ decay
using the CALO5 photon recombination scheme\cite{bib:lep4f}. The quoted
differential cross sections correspond to 
$\mathrm{d}[\sigma(\epem\rightarrow\WW\rightarrow\qqen) +
\sigma(\epem\rightarrow\WW\rightarrow\qqmn)]/
\mathrm{d}\cos\theta_{\mathrm{W}^-}$ within the above generator level
acceptance.

The differential cross section is measured in ten bins of $\costw$
with the data divided into four $\roots$ ranges: 
$180.0 - 185.0$\,GeV; $185.0 - 194.0$\,GeV;
$194.0 - 202.5$\,GeV; and $202.5 - 209.0$\,GeV.
Experimentally the angle $\costw$ can be obtained from the 
measured momenta of the two jets with the
lepton used to tag the charge of the W boson. However, to improve the
angular resolution a kinematic fit to the four momenta of the two
jets and the lepton is employed\cite{bib:opalmw}. If the fit
converges with a fit probability of $>0.1\,\%$\cite{bib:opalmw} 
the fitted jet momenta
are used. If the kinematic fit yields a fit probability of $<0.1\,\%$, 
which is the case for approximately 4\,\% of $\qqln$ events, 
$\costw$ is calculated from the measured jet four-momenta.
From MC the $\costw$ resolution is found to be approximately 0.05. 

The reconstructed $\costw$ distributions are corrected to the signal definition 
using the MC background estimates and a simple bin-by-bin efficiency 
correction. It has been verified that this simple 
bin-by-bin correction method is in good agreement with a more complete 
unfolding using the reconstructed to generator level migration.

The systematic uncertainties on the selection efficiencies
and background cross sections described above are propagated to
the differential cross section measurement. In addition it is
known from studies of lepton pair production at \LEPI\ that the
\OPAL\ MC underestimates the fraction of events where
the lepton track is assigned the wrong charge\cite{bib:zedometry}. 
This arises from imperfect tracking in the region of the jet chamber 
anode planes. For the data considered here the MC predicts that
$0.5\,\%$ of tracks are assigned the wrong charge. Based on 
previous studies\cite{bib:zedometry} it is estimated
that the corresponding number for data is $(1.0\pm0.5)\,\%$. 
In deriving the efficiency corrections, the MC reconstructed $\costw$ 
distributions are corrected for this difference and the full size of 
the correction is taken as the charge identification systematic uncertainty. 

\begin{figure}
 \begin{center}
   \epsfig{file=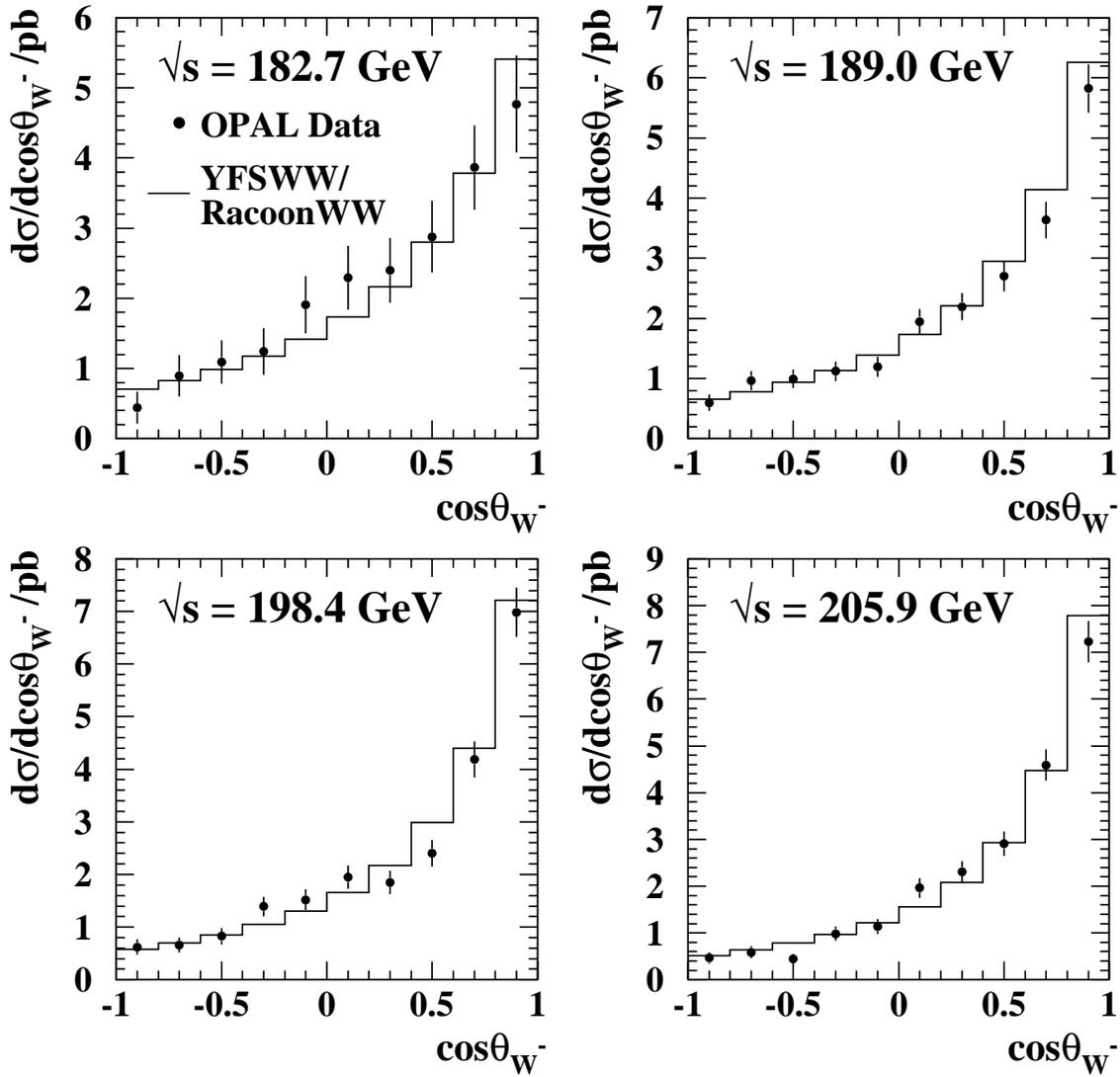,width=\textwidth}
   \caption{ 
     The measured W$^-$ polar angle differential cross section for 
     $\qqev$ and $\qqmv$ events within the acceptance defined in the text.
     The measurements are shown for the four energy bins described in
     the text. The measured cross sections (points) are compared to the
     theoretical expectations (histograms) from \YFSWW\ and \RACWW\
     (indistinguishable on this scale). 
     \label{fig:dsdcost}
     }
 \end{center}
\end{figure}

The measured differential cross sections in the 10 bins of
$\costw$ for the four energy ranges are shown in Figure~\ref{fig:dsdcost} and the 
results are given in Table~\ref{tab:costw}. 
The data are in good agreement with the SM expected generator level 
distributions obtained from either \YFSWW\ or \RACWW. 
Although the differential cross sections for these data have not been 
published previously, it should be noted that a deviation from the SM
would have shown up in the OPAL triple gauge coupling analysis\cite{bib:opaltgc}
which uses similar distributions.

 \begin{table}[htbp]
   \centering
    \begin{tabular}{c|cccc} \hline\hline
             & \multicolumn{4}{c}{Differential cross section [pb]}  \\ 
 \costw\ bin & $\langle\roots\rangle=182.7$\,GeV & $\langle\roots\rangle=189.0$\,GeV 
             & $\langle\roots\rangle=198.4$\,GeV & $\langle\roots\rangle=205.9$\,GeV \\ \hline 
 $-1.0\rightarrow-0.8$ &$0.44\pm0.22\pm0.02$ &$0.60\pm0.14\pm0.03$ &$0.62\pm0.15\pm0.04$ &$0.46\pm0.12\pm0.04$\\ 
 $-0.8\rightarrow-0.6$ &$0.90\pm0.30\pm0.02$ &$0.97\pm0.16\pm0.02$ &$0.66\pm0.15\pm0.02$ &$0.59\pm0.13\pm0.02$\\ 
 $-0.6\rightarrow-0.4$ &$1.09\pm0.31\pm0.01$ &$1.00\pm0.16\pm0.01$ &$0.83\pm0.15\pm0.01$ &$0.44\pm0.11\pm0.02$\\ 
 $-0.4\rightarrow-0.2$ &$1.24\pm0.33\pm0.01$ &$1.12\pm0.17\pm0.01$ &$1.39\pm0.19\pm0.01$ &$0.98\pm0.15\pm0.01$\\ 
 $-0.2\rightarrow\phantom{+}0.0$ 
             &$1.91\pm0.41\pm0.01$ &$1.19\pm0.17\pm0.01$ &$1.52\pm0.20\pm0.01$ &$1.14\pm0.16\pm0.01$\\ 
 $\phantom{+}0.0\rightarrow+0.2$ 
             &$2.29\pm0.45\pm0.01$ &$1.95\pm0.21\pm0.01$ &$1.95\pm0.22\pm0.01$ &$1.96\pm0.21\pm0.01$\\ 
 $+0.2\rightarrow+0.4$ &$2.40\pm0.46\pm0.01$ &$2.20\pm0.23\pm0.01$ &$1.85\pm0.22\pm0.01$ &$2.31\pm0.23\pm0.01$\\ 
 $+0.4\rightarrow+0.6$ &$2.88\pm0.51\pm0.02$ &$2.71\pm0.26\pm0.01$ &$2.41\pm0.25\pm0.01$ &$2.91\pm0.26\pm0.02$\\ 
 $+0.6\rightarrow+0.8$ &$3.87\pm0.60\pm0.02$ &$3.64\pm0.31\pm0.02$ &$4.19\pm0.34\pm0.03$ &$4.59\pm0.33\pm0.03$\\ 
 $+0.8\rightarrow+1.0$ &$4.77\pm0.69\pm0.03$ &$5.83\pm0.40\pm0.04$ &$6.98\pm0.47\pm0.04$ &$7.23\pm0.44\pm0.05$\\ 
 \hline\hline
 \end{tabular}
 \caption{\label{tab:costw} The measured differential cross section,
$\mathrm{d}[\sigma(\epem\rightarrow\WW\rightarrow\qqen) +
\sigma(\epem\rightarrow\WW\rightarrow\qqmn)]/
\mathrm{d}\cos\theta_{\mathrm{W}^-}$ expressed in ten bins of $\costw$ for
 the four centre-of-mass energy ranges. The cross sections correspond to the
 \CC\ set of diagrams with the additional requirement that the charged lepton 
 is more than $20^\circ$ from the beam axis, $20^\circ<\theta_{\ell^\pm}<160^\circ$.
 For each entry, the first uncertainty is statistical and the second systematic. 
 }
 \end{table}

\section{Conclusions}

From a total data sample of \lumi~pb$^{-1}$ recorded with 
$\epem$ centre-of-mass energies of $\roots = 161-209$~GeV 
with the OPAL detector at LEP \nsel\ W-pair 
candidate events are selected. The combined data samples is almost a factor
three larger than the previous OPAL publication. 
This large sample of events
has enabled a significant reduction in a number of systematic
uncertainties compared with our previous publications.

The data are used to test the SM description of $\WW$ production in the
centre-of-mass range $\roots = 161-209$~GeV. The W-pair production cross sections 
at 10 different centre-of-mass energies are found to be consistent
with the Standard Model expectation:
$$ \mathrm{data}/\mathrm{SM} = \rsmresult.$$
The data are then used to determine the W boson leptonic branching fractions:
\begin{eqnarray*}
  \Br(\Wtoen) & = &  \brevresult  \\ 
  \Br(\Wtomn) & = &  \brmvresult  \\ 
  \Br(\Wtotn) & = &  \brtvresult.
\end{eqnarray*}
These results are consistent with lepton universality of the
charged current weak interaction and with the results of the other LEP 
collaborations\cite{bib:xsaleph, bib:xsdelphi, bib:xsl3}.
Assuming lepton universality, the branching ratio to hadrons is determined to be \brqqresult\
from which the CKM matrix element \Vcs\ is determined
to be \vcsresult. The differential cross section as a function
of the W$^-$ production angle is measured for the $\Sqqev$ and
$\Sqqmv$ final states and found to be consistent with the SM 
expectation.

\bigskip
\noindent
{\large\bf Acknowledgements}

\par
\noindent
We particularly wish to thank the SL Division for the efficient operation
of the LEP accelerator at all energies
 and for their close cooperation with
our experimental group.  In addition to the support staff at our own
institutions we are pleased to acknowledge the  \\
Department of Energy, USA, \\
National Science Foundation, USA, \\
Particle Physics and Astronomy Research Council, UK, \\
Natural Sciences and Engineering Research Council, Canada, \\
Israel Science Foundation, administered by the Israel
Academy of Science and Humanities, \\
Benoziyo Center for High Energy Physics,\\
Japanese Ministry of Education, Culture, Sports, Science and
Technology (MEXT) and a grant under the MEXT International
Science Research Program,\\
Japanese Society for the Promotion of Science (JSPS),\\
German Israeli Bi-national Science Foundation (GIF), \\
Bundesministerium f\"ur Bildung und Forschung, Germany, \\
National Research Council of Canada, \\
Hungarian Foundation for Scientific Research, OTKA T-038240, 
and T-042864,\\
The NWO/NATO Fund for Scientific Research, the Netherlands.\\

\newpage

\end{document}